\documentclass[8.5pt]{article}
\usepackage[super,sort&compress,comma]{natbib} 
\usepackage{mhchem}
\usepackage{times,mathptmx}
\usepackage{sectsty}
\usepackage{balance} 

\usepackage{graphicx} 
\usepackage{lastpage}
\usepackage[format=plain,justification=raggedright,singlelinecheck=false,font=small,labelfont=bf,labelsep=space]{caption} 
\usepackage{fancyhdr}
\usepackage{setspace}
\usepackage{lineno}
\usepackage{amssymb}

\begin{document}

\thispagestyle{plain}
\fancypagestyle{plain}{
\renewcommand{\headrulewidth}{1pt}}
\renewcommand{\thefootnote}{\fnsymbol{footnote}}
\renewcommand\footnoterule{\vspace*{1pt} 
\hrule width 3.4in height 0.4pt \vspace*{5pt}} 
\setcounter{secnumdepth}{5}

\makeatletter 
\def\subsubsection{\@startsection{subsubsection}{3}{10pt}{-1.25ex plus -1ex minus -.1ex}{0ex plus 0ex}{\normalsize\bf}} 
\def\paragraph{\@startsection{paragraph}{4}{10pt}{-1.25ex plus -1ex minus -.1ex}{0ex plus 0ex}{\normalsize\textit}} 
\renewcommand\@biblabel[1]{#1}            
\renewcommand\@makefntext[1] 
{\noindent\makebox[0pt][r]{\@thefnmark\,}#1}
\makeatother 
\renewcommand{\figurename}{\small{Fig.}~}
\sectionfont{\large}
\subsectionfont{\normalsize} 

\fancyfoot{}
\fancyfoot[RO]{\footnotesize{\sffamily{1--\pageref{LastPage} ~\textbar  \hspace{2pt}\thepage}}}
\fancyfoot[LE]{\footnotesize{\sffamily{\thepage~\textbar\hspace{3.45cm} 1--\pageref{LastPage}}}}
\fancyhead{}
\renewcommand{\headrulewidth}{1pt} 
\renewcommand{\footrulewidth}{1pt}
\setlength{\arrayrulewidth}{1pt}
\setlength{\columnsep}{6.5mm}
\setlength\bibsep{1pt}

  \begin{@twocolumnfalse}
\noindent\LARGE{\textbf{An equilibrium double-twist model for the radial structure of collagen fibrils}}
\vspace{0.6cm}

\noindent\large{\textbf{Aidan I Brown, Laurent Kreplak\textit{$^{a, \ast}$}, and Andrew D Rutenberg\textit{$^{b, \ast}$}}}\vspace{0.5cm}
\footnotetext{{Department of Physics and Atmospheric Science, Dalhousie University, Halifax, NS, Canada, B3H 4R2}}
\footnotetext{\textit{$^{a}$~E-mail: kreplak@dal.ca}}
\footnotetext{\textit{$^{b}$~E-mail: andrew.rutenberg@dal.ca}}

\noindent\textit{\small{\textbf{Received Xth XXXXXXXXXX 20XX, Accepted Xth XXXXXXXXX 20XX\newline
First published on the web Xth XXXXXXXXXX 200X}}}
\noindent \textbf{\small{DOI: 10.1039/b000000x}}
\vspace{0.6cm}

\noindent \normalsize{ 
Mammalian tissues contain networks and ordered arrays of collagen fibrils originating from the periodic self-assembly of helical 300 nm long tropocollagen complexes. The fibril radius is typically between 25 to 250 nm, and tropocollagen at the surface appears to exhibit a characteristic twist-angle with respect to the fibril axis. Similar fibril radii and twist-angles at the surface are observed \emph{in vitro}, suggesting  that these features are controlled by a similar self-assembly process. In this work, we propose a physical mechanism of equilibrium radius control for collagen fibrils based on a radially varying double-twist alignment of tropocollagen within a collagen fibril. The free-energy of alignment is similar to that of liquid crystalline blue phases, and we employ an analytic Euler-Lagrange and numerical free energy minimization to determine the twist-angle between the molecular axis and the fibril axis along the radial direction.  Competition between the different elastic energy components, together with a surface energy,  determines the equilibrium radius and twist-angle at the fibril surface. A simplified model with a twist-angle that is linear with radius is a reasonable approximation in some parameter regimes, and explains a  power-law dependence of radius and twist-angle at the surface as parameters are varied.  Fibril radius and twist-angle at the surface corresponding to an equilibrium free-energy minimum are consistent with existing experimental measurements of collagen fibrils.  Remarkably, in the experimental regime, all of our model parameters are important for controlling equilibrium structural parameters of collagen fibrils. 
}\vspace{0.5cm} \end{@twocolumnfalse} 

\section{Introduction}
Collagen is the most abundant protein in mammalian tissues, providing mechanical strength to tissues such as bone, tendon, ligament, and skin.  Seven (I, II, III, V, XI, XXIV, and XXVII) of the 28 reported varieties of collagen form fibrils \cite{exposito10}. The spatial organization of  fibrils and their radii are characteristic of each tissue type \cite{parry84}, and {\em in vivo} the fibril radius changes with both age and loading history \cite{parry78, patterson-kane97}.  \emph{In vitro}, the fibril radius depends on assembly conditions \cite{mcpherson85, mosser06} such as collagen concentration, pH, and ionic strength, as well as on the type of collagen(s) present in the fibril \cite{birk90}.  

At the molecular level collagen fibrils are linear aggregates of $\sim$300 nm long, $\sim$1 nm wide tropocollagen complexes with a distinctive triple-helical structure \cite{hulmes02, hall58}. Both full-length  tropocollagen  and sonicated fragments form cholesteric phases \emph{in vitro} at high protein concentrations, above 800 mg/ml \cite{mosser06, giraud-guille92, giraud-guille89}. Cholesteric pitch and other aspects of collagen liquid crystallinity have been reviewed in detail \cite{giraudguillle08}. The measured cholesteric pitch varies between 0.5 and 2 $\mu$m depending on experimental conditions \cite{mosser06}.  At lower concentration and \emph{in vivo}, tropocollagen complexes pack laterally in a semi-crystalline fashion to form 20 to 500 nm diameter fibrils \cite{mosser06, hulmes02}. The details of the lateral packing of tropocollagen complexes within a fibril remain unclear, but accepted packing models have approximately a local hexagonal structure with a concentric superstructure \cite{hulmes02, hulmes95} and roughly $4000$ tropocollagen complexes are needed per 100 nm of fibril diameter \cite{hulmes95}. 

The axis of the tropocollagen complexes does not lay perfectly parallel to the fibril axis: X-ray scattering images of tendons displays arcs along the axis of fibrils with an opening angle of roughly 15$^{\circ}$  \cite{doucet11}, and fibrils imaged by electron microscopy (EM) show twisted morphologies with angular mismatch between the molecular and fibrillar axes of up to 20$^{\circ}$ at their surface \cite{holmes01, raspanti89, lillie77}.  Consistent with these measurements, Bouligand \emph{et al} \cite{bouligand85} described a double twist configuration in EM of reconstituted collagen fibrils. The same double twist configuration was also proposed by Hukins \emph{et al} to explain changes in the X-ray scattering of drying elastoidin spicules \cite{hukins76}. Double twist configurations are used to explain liquid crystal blue phases that occur near the isotropic to cholesteric transition for small chiral molecules with a small cholesteric pitch \cite{meiboom81, meiboom83}. In a cholesteric phase the director field rotates along one preferential direction, whereas in a double twist configuration the molecular orientation depends on a radial coordinate in a cylindrical domain \cite{degennes95, rey10} (see Fig.~1, below).  In most models of blue phases these cylinders then form lattices, with isotropic phase in the gaps between tubes \cite{degennes95}.

Several models of collagen fibrils incorporate tilted tropocollagen molecules in a cylindrical geometry. The simplest is a  constant twist-angle fibril model \cite{holmes01}, where all molecules have the same twist-angle (orientation) with respect to the cylindrical axis. Recently, a two-phase model has been proposed \cite{raspanti11} with an axial core and a constant twist-angle sheath outside of the core. Closer to the blue phase models, a constant gradient of the twist-angle fibril model \cite{raspanti89} has molecules parallel to the axis at the fibril centre, with the twist-angle increasing linearly until the fibril surface is reached.  All of these models have been proposed to qualitatively reconcile EM images of intact and sectioned collagen fibrils.  However, they do not consider the energetics of the proposed configurations and so cannot address whether they reflect possible equilibrium states.

We model the collagen fibril as a cylindrical double twist configuration. A Frank free energy \cite{degennes95} is used to  describe the free energy per fibril volume, in conjunction with surface energy.  Euler-Lagrange equations are developed to minimize ``bulk'' energetics and then surface terms are added before numerically identifying global free-energy minima.  We explore the effects of  elastic constants associated with splay ($K_1$), twist ($K_2$),  bend ($K_3$), and saddle-splay ($K_{24}$) deformations of the director field, as well as inverse cholesteric pitch ($q_0$) and surface tension ($\gamma$). Notably, equal moduli ($K_1=K_2=K_3$ \cite{degennes95, chaikin95}) are not assumed. Rather, consistent with the large aspect ratio of tropocollagen, we allow $K_3/K_2$ to be as large as $30$ \cite{lee90, odijk86}.  We also investigate the specific roles of the inverse cholesteric pitch $q_0$,  the surface tension $\gamma$, and the saddle-splay modulus $K_{24}$. The model leads to collagen fibril surface twist-angle vs.\ radius relationships that are consistent with available experimental data. Accordingly, we propose that  equilibrium free-energy minimization controls the radius and twist-angles of many collagen fibrils. 

\begin{figure}[t]
\centering
  \begin{tabular}{cc}
    \includegraphics[width=2.0in]{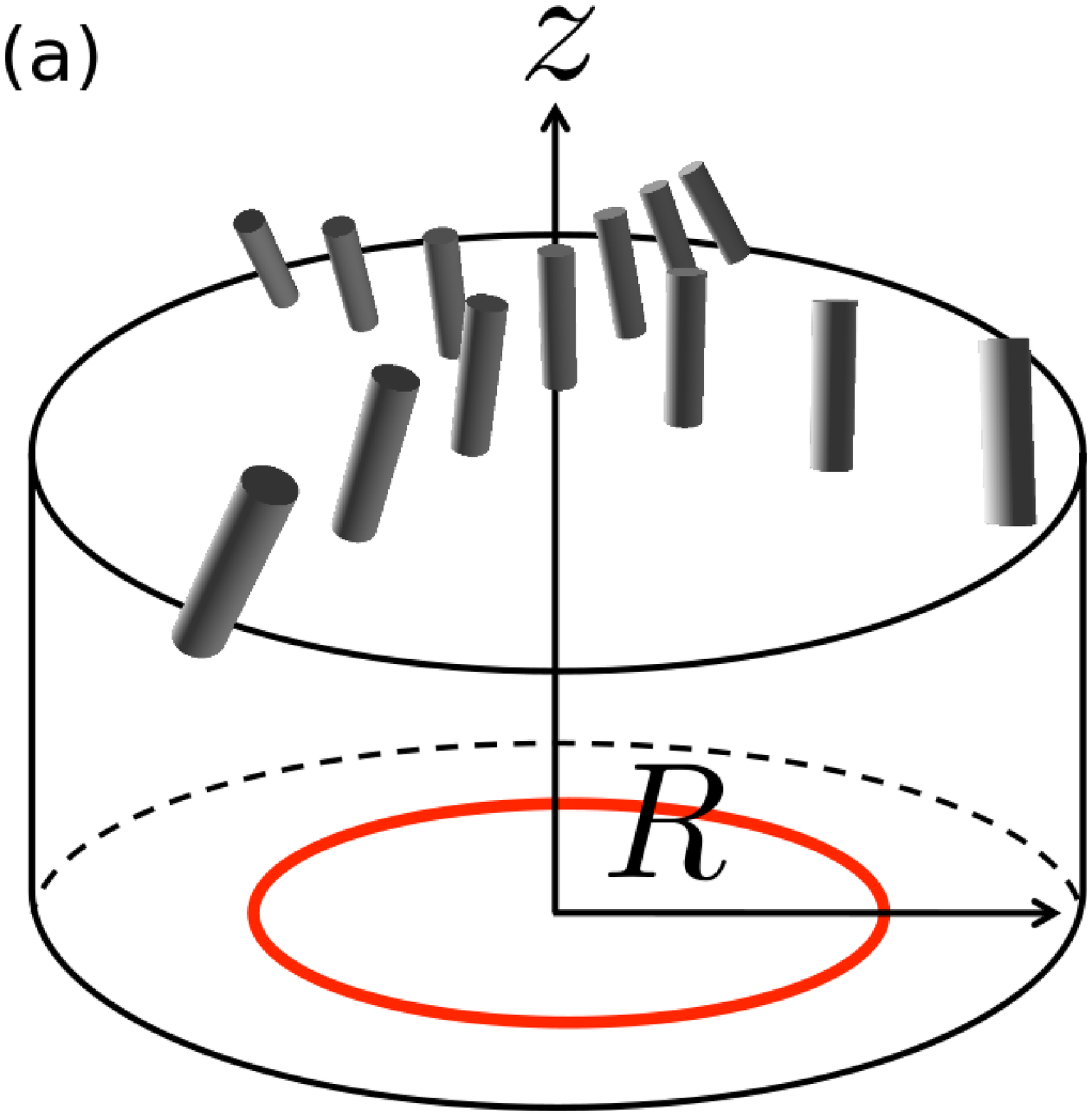} &\includegraphics[width=2.5in]{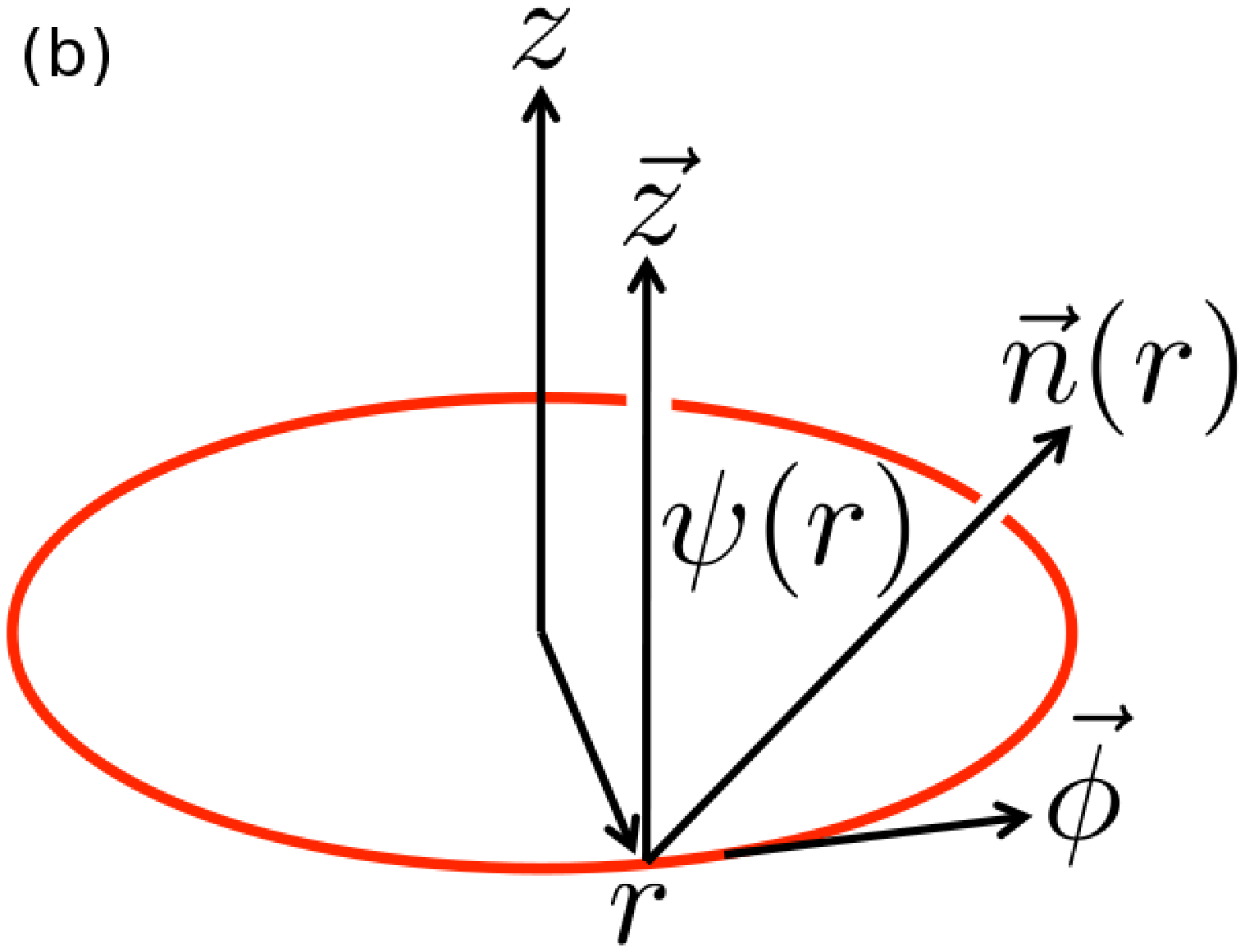}\\
  \end{tabular}
  \caption{(a) A double-twist configuration of tropocollagen molecules is schematically illustrated by tilted molecules within a fibril of radius $R$. The fibril axis is along the $z$ direction, as indicated. The tropocollagen director is axial at the centre of the fibril, but tilts away from the axis more as the radial distance increases. The thicker red outline indicates a radius $r<R$ within the fibril.  (b) We use cylindrical coordinates to describe the double-twist configuration inside a fibril.  For $r<R$, the twist-angle $\psi(r)$ only depends on the radius $r$. The local director $\vec{n}$, tangent $\vec{\phi}$, and axis $\vec{z}$ directions then are as illustrated. The director has no radial component, so that $n_{\phi}=-\sin\psi(r)$ and $n_{z}=\cos\psi(r)$.}
\label{fig:figure0}
\end{figure}

\section{Model}
\subsection{Frank free energy}
The Frank free energy density for a defect-free region of cholesteric-like liquid crystal with a spatially varying orientation vector $\vec{n}$ is \cite{degennes95} 
\begin{equation}
\label{eq:frank}
	f=\frac{K_1}{2}(\nabla\cdot\vec{n})^2+\frac{K_2}{2}(\vec{n}\cdot\nabla\times\vec{n}+q_0)^2+\frac{K_3}{2}(\vec{n}\times\nabla \times\vec{n})^2-K_{24}\nabla\cdot(\vec{n}\cdot\nabla\cdot\vec{n}+\vec{n}\times\nabla\times\vec{n}),
\end{equation}
where $K_1$, $K_2$, $K_3$, and $K_{24}$ terms correspond to the splay, twist, bend, and saddle-splay elastic energies, respectively.  In a cholesteric phase \cite{degennes95}, if the $z$ direction is chosen perpendicular to the plane of the liquid crystal layers, the director $\vec{n}$ in each layer will rotate along $z$: $n_x=\cos(q_0z)$, $n_y=\sin(q_0z)$, and $n_z=0$. Here $q_0$ is the inverse cholesteric pitch and quantifies the pitch of the helix,  $P=2\pi/q_0$. It is straightforward to show that $f_{cholesteric}=0$, so that thermodynamically stable phases other than the cholesteric must have the spatially-averaged $\langle f \rangle<0$.  The saddle-splay term ($K_{24}$) can be negative and so is necessary to achieve a stable blue phase \cite{meiboom81}.

\subsection{Fibril energy}
Our starting point is the common assumption for both blue phases \cite{degennes95, wright89, xing08} and collagen fibrils \cite{raspanti11, bouligand85, holmes01} that there is no radial component to the orientation vector. This is only strictly necessary at $r=0$ to avoid singularities in $f$, but it is assumed throughout the fibril. Assuming approximate homogeneity of the orientation field along the axial direction of a cylindrical fibril, the director may then be parameterized by a single twist-angle as shown in Fig.~\ref{fig:figure0}: $n_r=0$, $n_{\phi}=-\sin\psi(r)$, and $n_{z}=\cos\psi(r)$. The resulting free energy density for a fibril is then \cite{xing08}
\begin{equation}
	f_{fibril} = \frac{1}{2}K_2\left(q_0-\psi'-\frac{1}{r}\sin\psi\cos\psi\right)^2+\frac{1}{2}K_3\frac{\sin^4\psi}{r^2}-	\frac{K_{24}}{r}\frac{d	\sin^2\psi}{dr}, \label{eq:frankparameter}
\end{equation}
where the $\psi' \equiv d\psi/dr$. We identify the contributions due to $K_2$, $K_3$, and $K_{24}$, with $f_2$, $f_3$, and $f_{24}$, respectively. With a given fibril radius $R$, we can integrate $f_{fibril}$ to obtain the free energy per unit length due to the double-twist configuration: 
\begin{equation}
	\tilde{E}_{bulk} \equiv \int_0^R2\pi rfdr\equiv \tilde{E}_2+\tilde{E}_3+\tilde{E}_{24},
\end{equation}
where each $\tilde{E}_i$ is the contribution from the respective $f_i$. For the saddle-splay term we have exactly $\tilde{E}_{24}=-2\pi K_{24}\sin^2\psi(R)$.

A given cross-section $A$ of collagen fibre may be distributed between $N=A/(\pi R^2)$ individual fibrils which will each have a surface area per unit length of $2\pi R$. For a surface tension $\gamma$ this leads to an additional surface energy $E_{surf}=2\pi\gamma R$. For a given $A$, the total free energy per unit length is then
\begin{eqnarray}
	\tilde{E} &=& N(\tilde{E}_{bulk}+\tilde{E}_{surf})\\
	 &=& \frac{A}{\pi R^2}\left[\tilde{E}_2+\tilde{E}_3-2\pi K_{24}\sin^2\psi(R)+2\pi\gamma R\right].
\end{eqnarray}
This then directly gives us the configurational free-energy per unit volume of fibril
\begin{equation}
\label{eq:energy}
	E \equiv \frac{\tilde{E}}{A}=\frac{\tilde{E}_2}{\pi R^2}+\frac{\tilde{E}_3}{\pi R^2}-\frac{2K_{24}\sin^2\psi(R)}{R^2}+\frac{2\gamma}{R}.
\end{equation}
We will call $E$ the ``energy'' throughout the paper, and this quantity will be minimized to determine the equilibrium (minimal free-energy per unit volume) configuration for a collection of collagen fibrils. We note that $E_{cholesteric} = 0$. 

\subsection{Determining the director} \label{subsec:director}
While the twist-angle $\psi(r)$ is often approximated as having a constant gradient   \cite{degennes95,meiboom81,meiboom83,wright89}, we may determine it numerically without this approximation using a standard variational approach to extremize the bulk energy $E_2+E_3$. Assuming that arbitrary but small variations $\eta(r)$ in the twist-angle $\psi(r)$ do not change the bulk energy we obtain
\begin{equation}
\label{eq:elfirst}
	\int_0^R\left[r\frac{\partial f}{\partial\psi}-\frac{d}{dr}\left(r\frac{\partial f}{\partial\psi'}\right)\right]\eta(r)dr + \left[\eta(r)r			\frac{\partial f}{\partial\psi'}\right]_0^R=0.
\end{equation}
With no radial component to the director, where $\psi(0)=0$, we must have $\eta(0)=0$. However we cannot constrain $\eta(R)$, and requiring the second term of Eqn.~\ref{eq:elfirst} to independently vanish  gives us 
\begin{equation} 
	\label{eq:condition}
	K_2\left[q_0-\psi'(R)-\frac{\sin(2\psi(R))}{2R}\right]+\frac{K_{24}}{R}\sin(2\psi(R))=0.
\end{equation}
Similarly, arbitrary values of $\eta(r)$ force the integral in the first term in Eqn.~\ref{eq:elfirst} to vanish which gives us the Euler-Lagrange (E-L) equation
\begin{equation}
	r\frac{\partial(f_2+f_3)}{\partial\psi}=\frac{d}{dr}\left[r\frac{\partial f_2}{\partial\psi'}\right].
\end{equation}
Applying our previous expressions for $f_2$ and $f_3$ then gives us
\begin{equation}
	\label{eq:aftereulerlagrange}
	\psi'+r\psi''=q_0+\frac{\hat{K}^{-1}}{r}\sin^2\psi\sin(2\psi)-\cos(2\psi)\left[q_0-\frac{\sin(2\psi)}{2r}\right],
\end{equation}
where $\hat{K}^{-1}=K_3/K_2$.

\subsection{Numerical method}
Twist-angles $\psi(r)$ that satisfy Eqns.\ \ref{eq:condition} and \ref{eq:aftereulerlagrange} minimize the free-energy for a given fibril radius $R$. We  determine $\psi(r)$ numerically, using a modified midpoint method to solve Eqn.~\ref{eq:aftereulerlagrange} for a given $\psi'(0)$. The initial twist-angle gradient $\psi'(0)$ is then varied until the E-L solution also satisfies Eqn.~\ref{eq:condition}.  We check that the E-L solutions represent local minima of the free-energy with respect to $\psi'(0)$. These solutions determine the optimal twist-angle configuration for a given $R$. We then calculate the elastic and surface energies, and use Eqn.~\ref{eq:energy} to find $E(R)$. As illustrated in Fig.~S1, we check that the total energy represents a local minimum with respect to radius with $d^2E/dr^2>0$.

Since we have a largely numerical approach, we cannot definitively say that our solutions represent the global (as opposed to local) minimization of the free-energy. However, as illustrated in Fig.~S2, we have considered $E(R)$ for various parameterizations and we only ever find one minimum appropriate for collagen fibrils with radius $R< 1\mu m$ that is stable with respect to the cholesteric with $E<0$.

\subsection{Parameter values}  \label{subsec:parameters}
Unless otherwise stated, we will explore our model around default parameter values $K_2=10$pN, $K_3=300$pN, $K_{24}=K_2=10$pN, $\gamma=3$pN/$\mu$m, and $q_0 = \pi \mu m^{-1}$. Many of these parameters have not yet been directly measured for collagen, so, as described here, we rely on the liquid crystal literature for approximate elastic constants of solutions of long and/or chiral molecules. 

The twist modulus, $K_2$, has been estimated to be $\simeq10^{-6}$ dyne = 10pN for a dilute solution of chiral molecules in a conventional nematic \cite{degennes95}, a comparable value of  $3\times10^{-12}$N = 3pN  \cite{degennes95, meiboom81, chaikin95} has been used for blue phases. We do not expect $K_2$ to be significantly affected by the large aspect ratio of tropocollagen because it is approximately unchanged as molecular weight is varied \cite{toriumi84, taratuta88}.  Accordingly, we use $K_2=10$pN as our default value. 

While the three moduli of the Frank free energy, Eqn.\ \ref{eq:frank}, are commonly taken to be equal, the bend modulus, $K_3$, is affected by the length of the liquid crystalline molecule.  For liquid crystals composed of semi-flexible polymers the $K_3/K_2$ ratio saturates to a constant value for polymers much longer than their persistence length, with the ratio controlled by the persistence length \cite{lee90, odijk86}. For example, for the semi-rigid macromolecule poly-$\gamma$-benzyl glutamate, the saturation ratio of $K_3/K_2=30$ is reached for aspect ratios between 50 and 100 \cite{lee90}.  As tropocollagen complexes are greater than 100$\times$ longer than their width \cite{hall58}, we take the ratio $K_3=30K_2$ as our default ratio. However, the persistence length of collagen, and thus $K_3$, is not independent of environment --- in particular choice of solvent can affect collagen persistence length \cite{lovelady13}. We investigate smaller values of $K_3$ below. 
 
While the saddle-splay modulus is typically neglected in a bulk cholesteric phase because it is equivalent to a surface term \cite{meiboom81, degennes95}, surface terms cannot be neglected for double-twist cylinders. The saddle-splay modulus has been estimated \cite{degennes95} to satisfy $K_{24}\simeq K_2$, and we use this to determine our default value of $K_{24}=K_2=10$pN. We vary $K_{24}$ below. 

The surface tension of blue phases has been estimated to range from $5\times10^{-4}$erg cm$^{-2}$ = 0.5pN/$\mu$m for azoxyphenetole to $2.3\times10^{-2}$erg cm$^{-2}$ = 23pN/$\mu$m for methoxybenzylidene-butylaniline \cite{meiboom83}. Prost and de Gennes \cite{degennes95} use a value $10^{-2}$erg cm$^{-2}$ = 10pN/$\mu$m, which is within this range. In our case, we consider a concentrated solution of collagen in water \cite{mosser06}. The surface tension for a fibril immersed in a concentrated aqueous collagen solution should be lower than the surface tension of the same fibril in pure water. The latter should be similar to the surface tension of high molecular weight poly(ethylene Glycol) in potassium phosphate, which reaches as low as 100pN/$\mu$m \cite{oliveira12}. As a starting point for this study we use a default surface tension of $\gamma=3$pN/$\mu$m -- this is within the broad range used for blue phases.  We explore the effects of different surface tensions below. We expect surface tension to depend upon local conditions, for example in different tissue types or with different collagen concentrations.

The pitch, $P \equiv 2 \pi/q_0$, of the cholesteric phase of a lyotropic mesogen depends on both ionic strength and concentration \cite{stanley05}.  Cholesteric phases of DNA  \emph{in vitro} and \emph{in vivo} exhibit pitches ranging between 50 nanometers and 5 micrometers \cite{livolant91}, with the smallest pitches at the highest concentration.  Measured cholesteric pitches of collagen also decrease with concentration and vary between 0.5 and 2$\mu$m \cite{mosser06}, so that $q_0 \in [\pi, 4\pi]$ $\mu m^{-1}$. We use $q_0 = \pi \mu m^{-1}$ as our default value, corresponding to a high concentration solution of collagen, but we also explore other values below. 

\section{Results}
\subsection{Linear approximation}
\label{subsec:linear}
While we expect the gradient of the twist-angle at the centre of the fibril, $\psi'(0)$, to be of the same order as $q_0$ \cite{xing08}, it is a common approximation \cite{meiboom81, meiboom83, degennes95, wright89} to additionally assume that the gradient of the twist-angle is constant throughout the fibril.  We call this the linear approximation, since then $\psi(r)= \psi' r$. If we additionally restrict our attention to regimes where the twist angle is small, i.e. $\psi(r) \ll 1$, then the linear (small-angle) approximation is straight-forward analytically. Using Eqn.~\ref{eq:frankparameter} in Eqn.~\ref{eq:energy}, together with the linear small-angle approximation, we obtain 
\begin{equation}
\label{eq:linearenergyfirst}
	E_{linear}=\frac{N\tilde{E}}{A}=\frac{\tilde{E}}{\pi R^2}=\left[\frac{K_2}{2}(q_0-2\psi')^2-2K_{24}(\psi')^2\right]+\frac{K_3}{4}		(\psi')^4R^2+\frac{2\gamma}{R}.
\end{equation}
Minimizing this energy per unit volume with respect to radius $R$ leads to
\begin{equation}
\label{eq:linearrfirst}
R_{lin}=\left(\frac{4\gamma}{K_3 \psi'^4}\right)^{1/3}.
\end{equation}
For convenience, we restrict our attention to  $K_{24}=K_2$, which leads to
\begin{eqnarray}
	\psi'_{lin} &=& \frac{K_2^3q_0^3}{2K_3\gamma^2}, \\
	R_{lin} &=& \frac{4K_3\gamma^3}{q_0^4K_2^4},	\label{eq:linearr} \\
	\psi_{linear}(R) &=& \psi'R=\frac{2\gamma}{q_0K_2}, \label{eq:linearpsi} \\
	E_{lin} &=& \frac{K_2q_0^2}{2}\left(1-\frac{K_2^3q_0^2}{2\gamma^2K_3}\right).	\label{eq:linearenergy}
\end{eqnarray}
With the linear and small $\psi$ approximations, the fibril phase is stable with respect to the cholesteric (with $E_{lin}<0$) when $K_2^3q_0^2>2\gamma^2K_3$. Larger $K_3$ and $\gamma$ values reduce the stability of the fibril phase, while larger $K_2$ and $q_0$ values increase the stability of the fibril phase.

Consider the self-consistency of the small angle and linear approximations.  From Eqn.~\ref{eq:linearpsi}, a small surface angle requires $\gamma \lesssim q_0 K_2/2$, which together with the stability condition from Eqn.~\ref{eq:linearenergy} requires that $K_3 < 2 K_2$. This condition is violated for our collagen parameterization since $K_3/K_2 \approx 30$. This means that the small-angle linear approximation gives an unstable fibril phase with respect to the cholesteric.  Even if we ignore the cholesteric phase, the linear approximation requires that the term ignored in Eqn.~\ref{eq:aftereulerlagrange} is small --- i.e. that  $R \psi'' \ll \psi'$. Using our linear solution, we obtain $R \psi''/\psi' = 2 K_3/K_2 \psi(R)^2$. For our default parameter values the right-side is $\approx 2$ --- i.e. not small. In summary, the linear approximation is uncontrolled for larger values of $K_3/K_2$. 

In addition to the small-angle linear approximation, we also consider the linear approximation alone -- without any small angle approximation. We do this numerically, by enforcing $\psi'(r)=\psi' r$ instead of Eqns.\ \ref{eq:condition} and \ref{eq:aftereulerlagrange}, and then minimizing $E$ with respect to $\psi'$ and $R$. As shown in our figures below, the  linear approximation results largely agree with the small-angle linear approximation results --- this indicates that disagreements with the full numerical free-energy minimization arise largely from the linear approximation alone rather than from the (analytically convenient) small-angle approximation. Nevertheless, we will show below that Eqns.~\ref{eq:linearr} and \ref{eq:linearpsi} still provide valuable insight into the qualitative scaling behaviour of our results as model parameters are varied. 

\subsection{Existence of a stable free-energy minimum vs.\ fibril radius $R$} 
\begin{figure}[t]
\centering
  \begin{tabular}{cc}
    \includegraphics[width=3.0in]{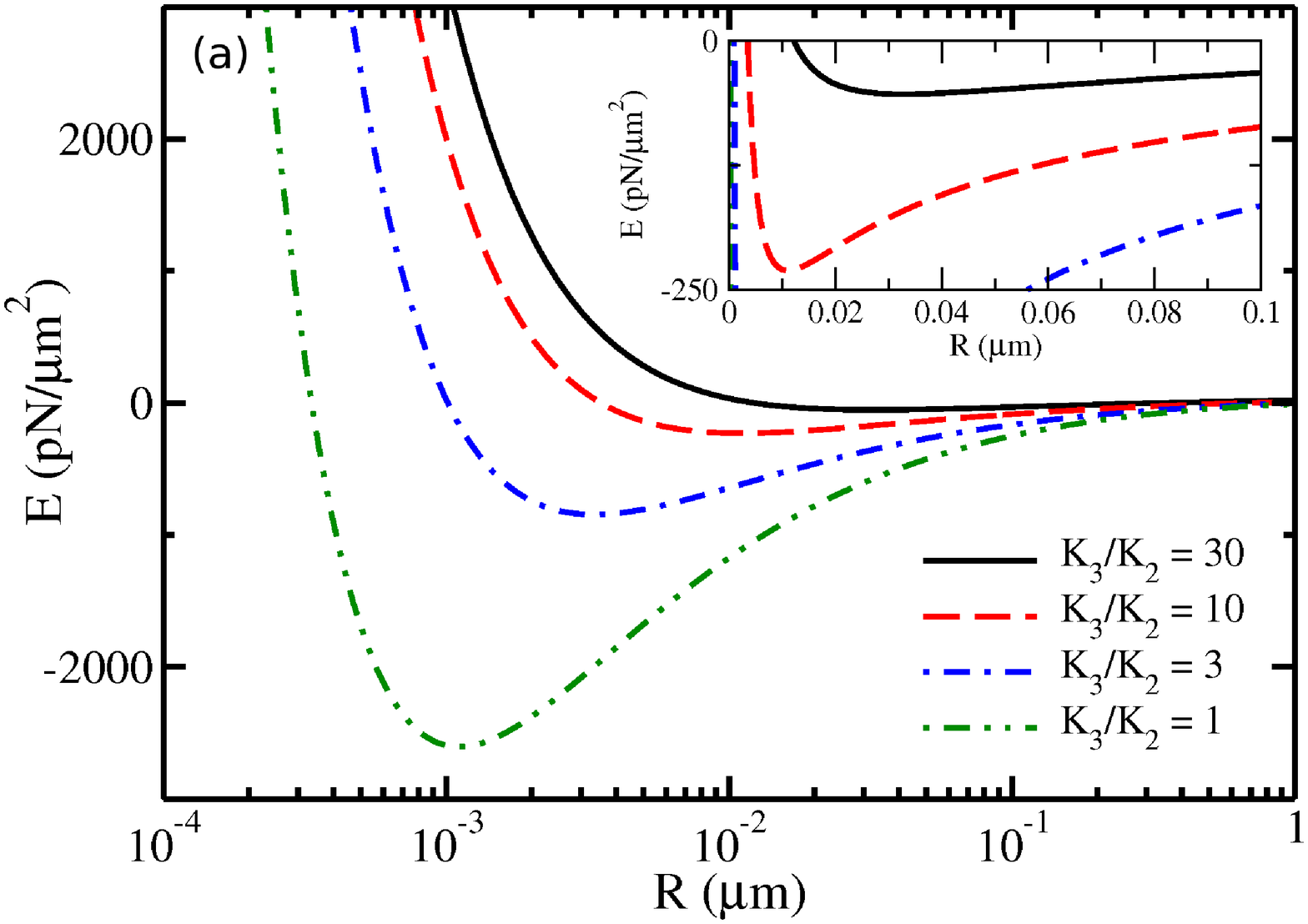} & \includegraphics[width=3.0in]{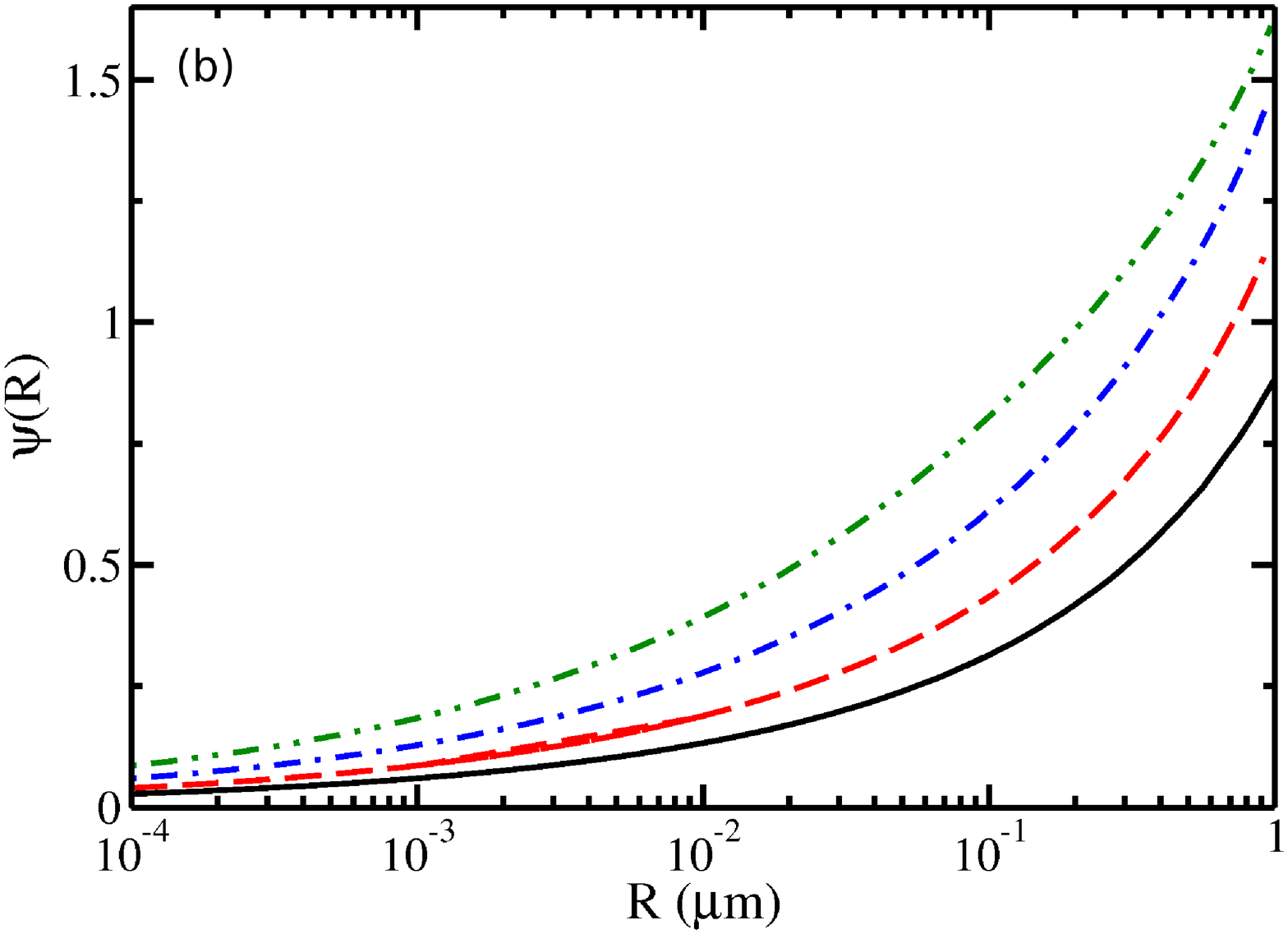}\\
    \includegraphics[width=3.0in]{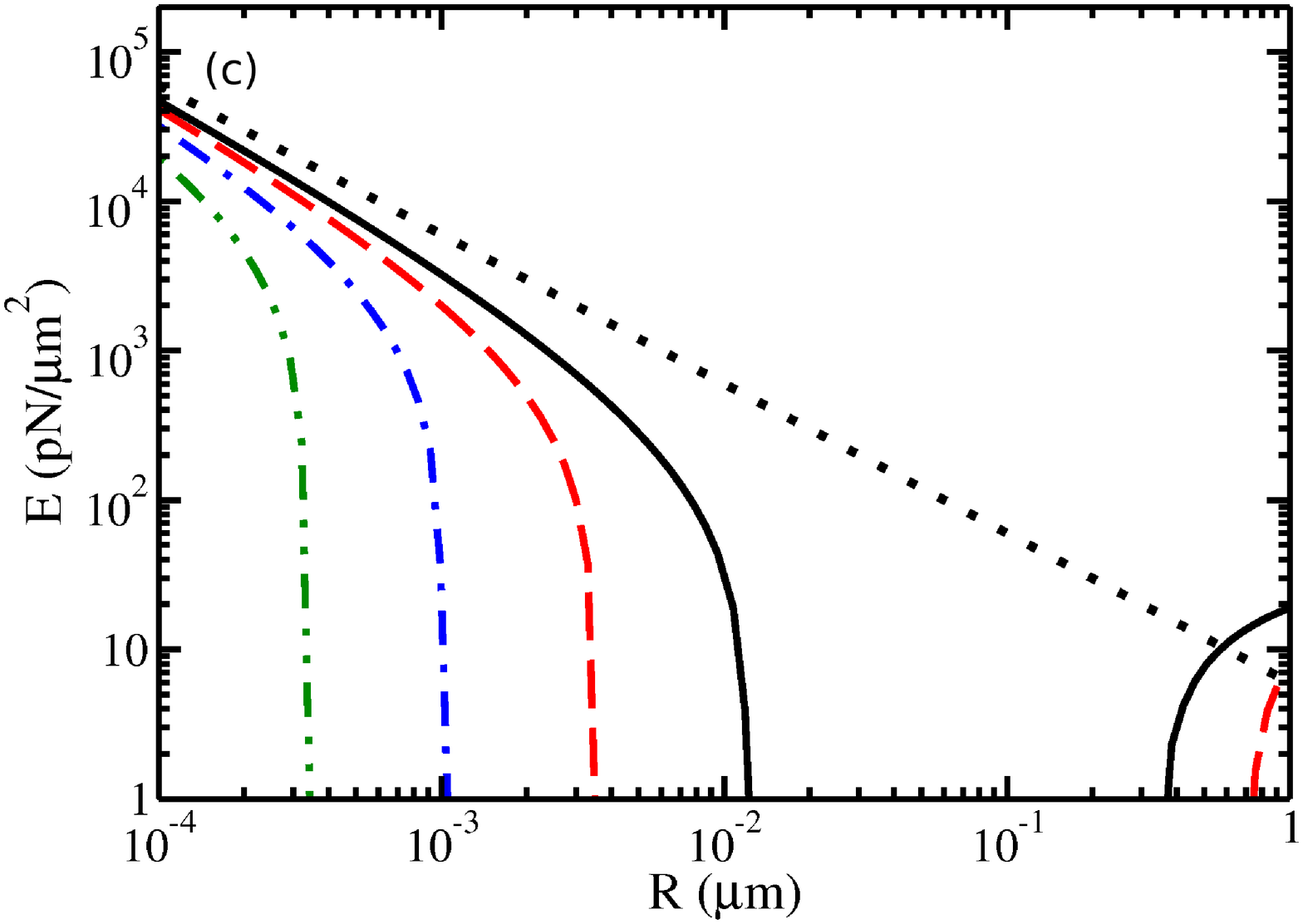} & \includegraphics[width=3.0in]{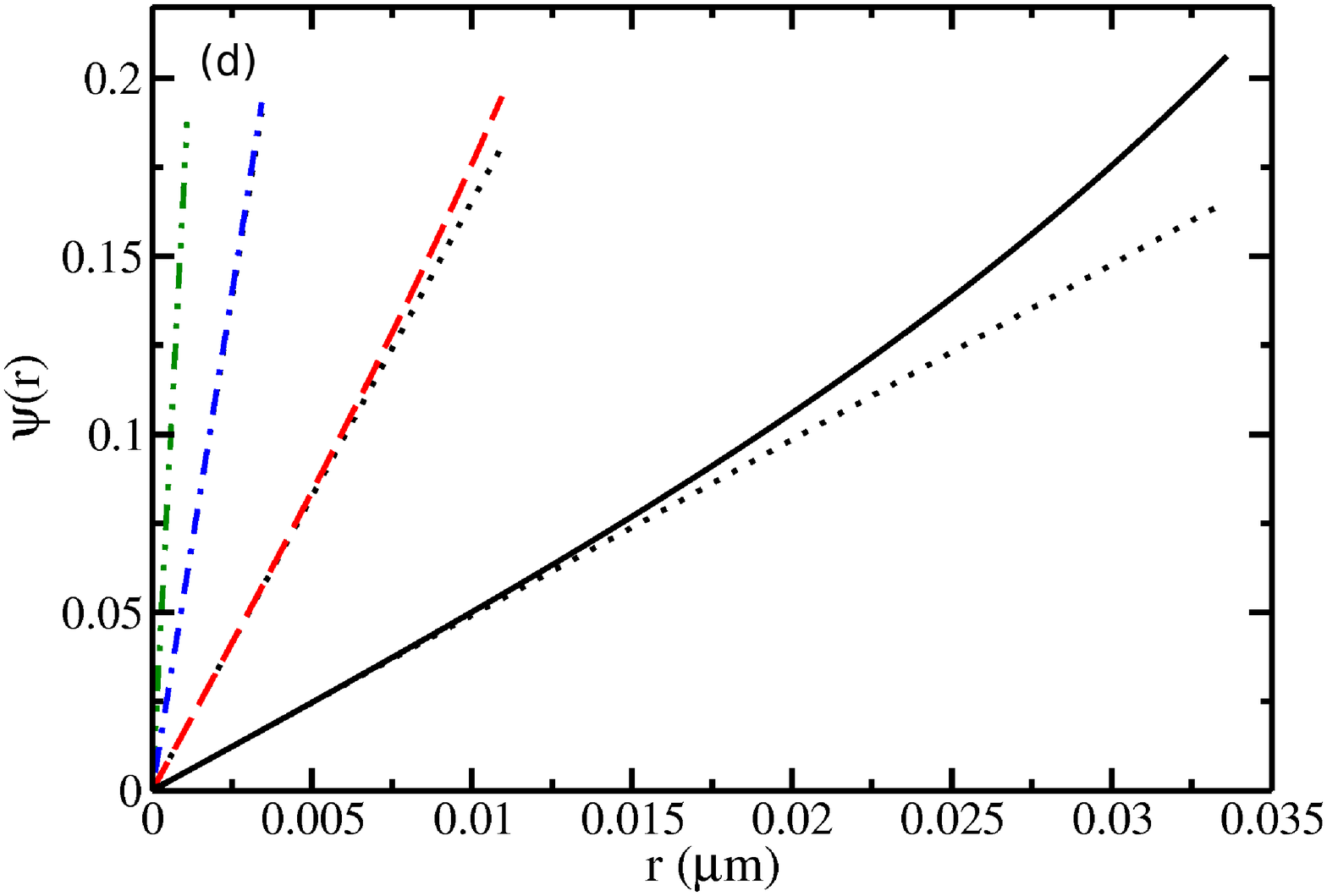}
  \end{tabular}
  \caption{(a) Energy per unit volume $E(R)$ vs. fibril radius $R$, for several $K_3/K_2$ ratios as indicated by the legend. $E(R)$ is given by Eqn.~\ref{eq:energy}, where the twist-angle satisfies Eqns.~\ref{eq:condition} and \ref{eq:aftereulerlagrange} to result in the minimal bulk free-energy at a given $R$.  Inset highlights the stable ($E<0$) minima for $K_3=30K_2$ and $K_3=10K_2$. (b) Corresponding twist-angle at the surface $\psi(R)$ vs $R$.  (c) Positive energies for the same data as (a). Dotted black line is the surface tension contribution. (d) Corresponding to the minima in (a), the twist-angle $\psi(r)$ vs.\ radial coordinate $r$ for $r \in [0,R]$.  The dashed black lines represent linear extrapolations using the twist-angle gradient at small $r$.  The legend in (a) applies to the entire figure. We use our default values of $q_0=\pi$ $\mu$m$^{-1}$,  $K_2=10$pN, $K_{24}=10$pN, and $\gamma=3$pN/$\mu$m.}
  \label{fig:figure1}
\end{figure}

We show in Fig.~\ref{fig:figure1}(a) the total energy per unit volume $E$, and in Fig.~\ref{fig:figure1}(b) the twist-angle at the surface $\psi(R)$, both as a function of fibril radius $R$, while $K_3$ is varied as indicated in the legend (and $K_2=10$ pN). While the equal constant approximation \cite{chaikin95, degennes95} typically assumes a bend modulus of $K_3=K_2$, the large length to width ratio of tropocollagen makes values up to $K_3 = 30 K_2$  appropriate \cite{lee90, odijk86}.  

Fig.~\ref{fig:figure1}(a) illustrates that there is a well-defined minimum in the energy vs.\ $R$, indicating that the radius of a collagen fibril can be controlled by equilibrium free-energy minimization.  See also Figs.~S1 and S2. This minimum is deeper and occurs at lower radii for smaller $K_3$. The inset shows that the minimum still has a negative free-energy at the largest $K_3$ explored, indicating thermodynamic stability vs.\ the cholesteric phase. We also note that at a given $R$, $E(R)$ monotonically increases with $K_3$ --- as expected since larger $K_3$ have larger positive contributions in Eqn.~\ref{eq:frank}.  Similarly, we see in Fig.~\ref{fig:figure1}(b) that $\psi(R)$ is larger for \emph{smaller} $K_3$ values --- as expected due to the lower energetics of bending. The positive region of the energies in Fig.~\ref{fig:figure1}(a) are plotted in Fig.~\ref{fig:figure1}(c). At low radii, the energy is dominated by the surface tension, which is shown in Fig.~\ref{fig:figure1}(c) as a dotted black line. The fibril energies trend towards the surface tension at low radii, keeping $E>0$ in this regime. 

The $\psi(r)$ corresponding to the energy minima of Fig.~\ref{fig:figure1}(a) are plotted in Fig.~\ref{fig:figure1}(d) for $r \leq R$. All the $K_3$ values reach similar twist-angle at the surface $\psi(R)$, where $r=R$, but for low $K_3$ values they do this at a much lower radius than higher $K_3$ values. $\psi(r)$ is close to linear for all four $K_3$ values shown in Fig.~\ref{fig:figure1}(d). The dotted black lines in Fig.~\ref{fig:figure1}(d) are the linear extrapolations of the initial slope of the numerical curves. For low $K_3$ values the linear and numerical curves are indistinguishable, but as $K_3$ increases the difference grows. The disagreement is despite the small values of the twist-angle, which reinforces our observation that the linear approximation is not self-consistent for larger values of $K_3$ --- and indicates that this is independent of any small-angle approximation.  

\begin{figure}[ht]
\centering
  \begin{tabular}{cc}
    \includegraphics[width=3.0in]{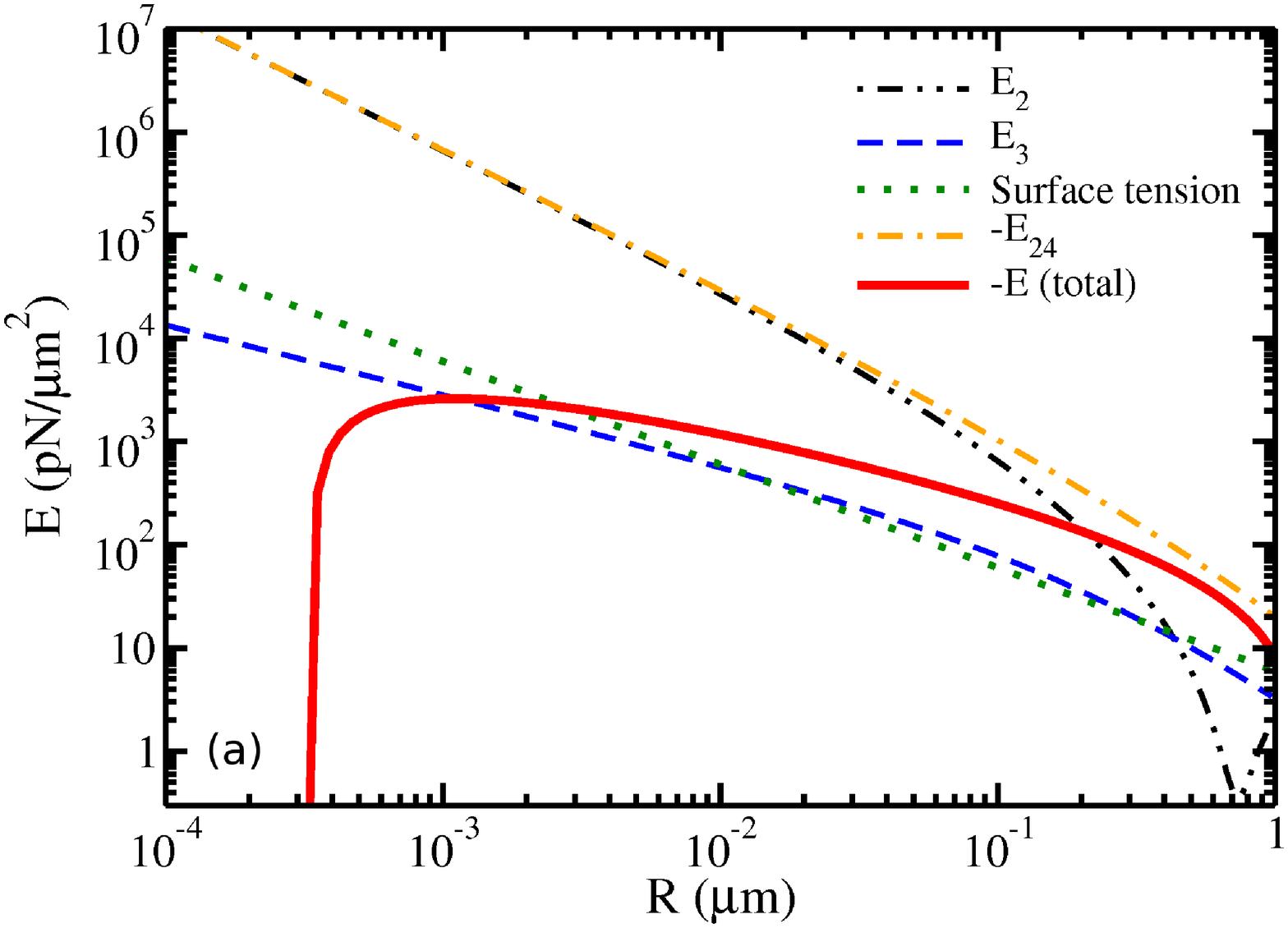} & \includegraphics[width=3.0in]{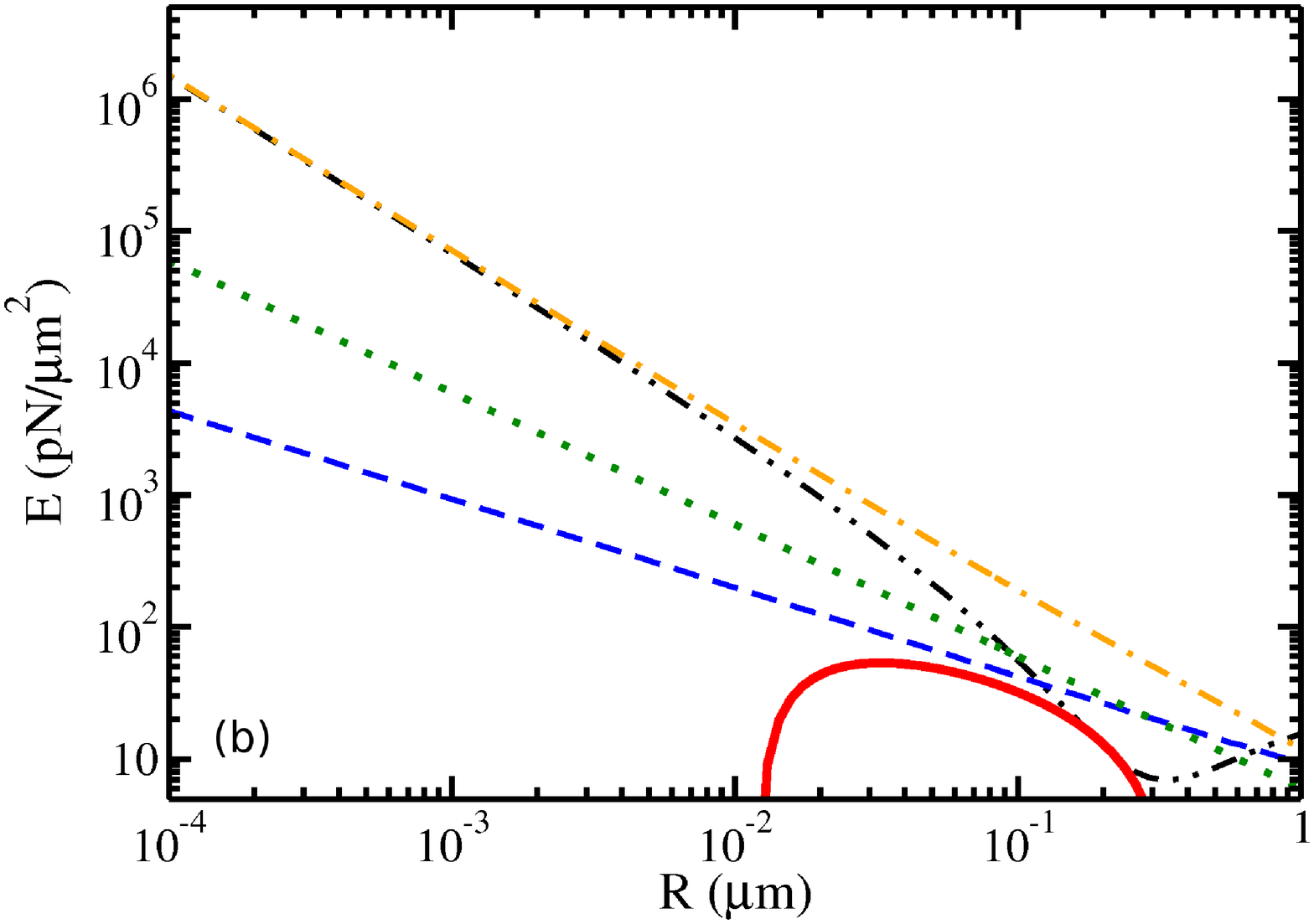}\\
  \end{tabular}
  \caption{Contributions of different elastic components to the total energy $E$ vs.\ $R$ for (a) $K_3/K_2=1$ and (b) $K_3/K_2=30$.  From Eqn.~\ref{eq:energy}, the twist term ($\tilde{E}_2/(\pi R^2)$, thin dashed-double-dotted black line) is $E_2$, bend term ($\tilde{E}_3/(\pi R^2)$, thin dashed blue line) is  $E_3$, saddle-splay term ($-2K_{24} \psi^2(R)/R^2$, thin dotted green line) is  $E_{24}$, and the surface tension term (thin dash-dotted orange line) is $2\gamma/R$. The sum of these individual terms is the total $E$ (thicker solid red line).  Note that negative $E_{24}$ and negative total $E$ are plotted, as indicated. The legend in (a) applies to both panels. We use the default values of $q_0=\pi$ $\mu$m$^{-1}$, $K_2=10$pN, $K_{24}=10$pN, and $\gamma=3$pN/$\mu$m.}
  \label{fig:figure2}
\end{figure}

Fig.~\ref{fig:figure2} shows the different components contributing to the total energy in Fig.~\ref{fig:figure1}(a), using the largest and smallest $K_3$ values in Fig.~\ref{fig:figure2}(a) and (b), respectively. The different energy components plotted in Fig.~\ref{fig:figure2} are the different terms on the right-hand-side of Eqn.~\ref{eq:energy}: twist $E_2$, bend $E_3$, saddle-splay $E_{24}$, and the surface tension are the first, second, third, and fourth terms respectively. The total energy $E$ is their sum. We have plotted negative $E_{24}$ and negative $E$, as indicated, so that we can use a log-log scale.

The plots for both $K_3$ values are qualitatively similar. At low $R$ the component terms with the largest magnitudes are $E_2$ and $E_{24}$ --- these are nearly equal in magnitude but have opposite sign. At low $R$, the next largest term is the surface tension, which makes the total energy positive. As $R$ increases the magnitude of $E_2$ drops more rapidly than $E_{24}$, which allows the total energy to become negative --- corresponding to stable fibrils.  At larger $R$, $E_2$ begins to {\em increase}, leading to a smaller magnitude of $E$ --- i.e. an energy minimum. We note that as $K_3$ is increased from $K_2$ to $30 K_2$, the range of $R$ that corresponds to stable fibrils with respect to the cholesteric phase (with $E<0$) decreases significantly. 

\subsection{Parameter variation}
\label{subsec:parameter}
For each parameter combination of inverse cholesteric pitch $q_0$, twist modulus $K_2$, bend modulus $K_3$, saddle-splay modulus $K_{24}$, and surface tension $\gamma$, we identify the  minimum of  $E(R)$ --- as illustrated in Fig.~\ref{fig:figure1}(a).  This gives  the equilibrium fibril radius $R$, twist-angle at the surface $\psi(R)$, and total energy $E$.  In this section, we show how those equilibrium values depend upon the model parameters. We note that fibril radius $R$ and twist-angle at the surface $\psi(R)$ are experimentally accessible, while $E$ must be negative to correspond to a stable phase with respect to the bulk cholesteric. 

\subsubsection{Inverse cholesteric pitch $q_0$ dependence\\}
\label{subsubsec:bend}
\begin{figure}[ht]
\centering
    \begin{tabular}{cc}
      \includegraphics[width=3.0in]{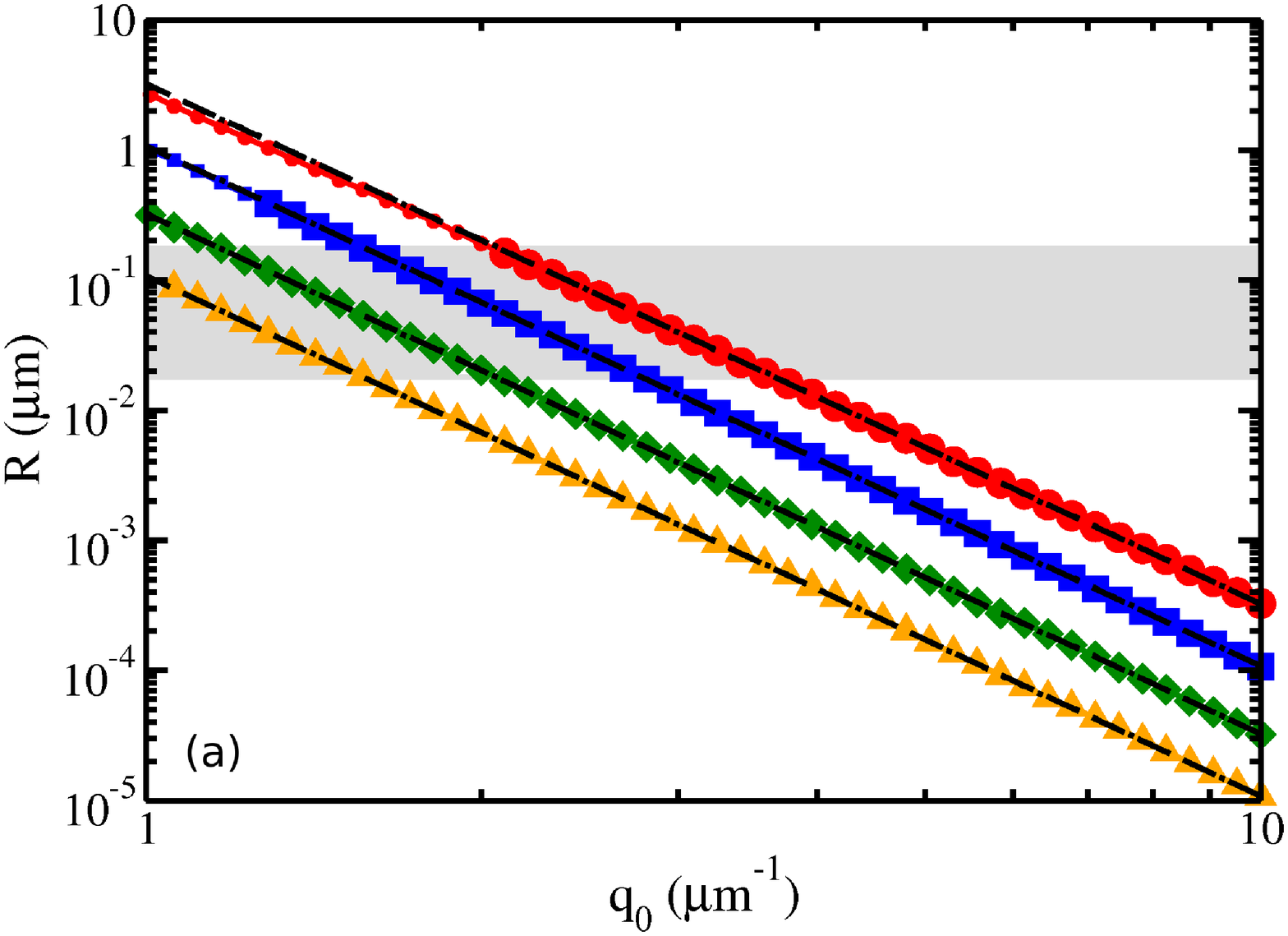} & \includegraphics[width=3.0in]{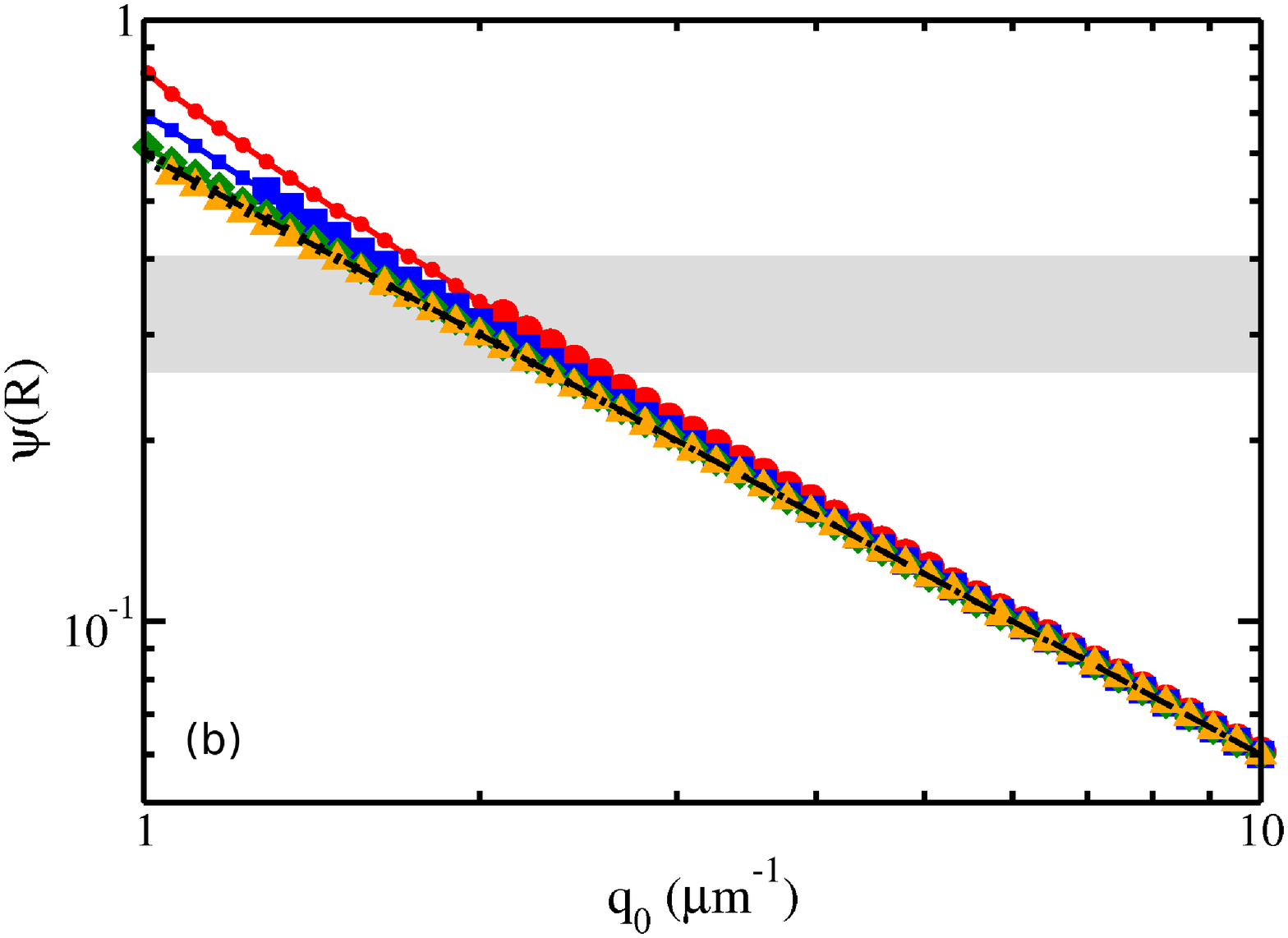}\\
      \includegraphics[width=3.0in]{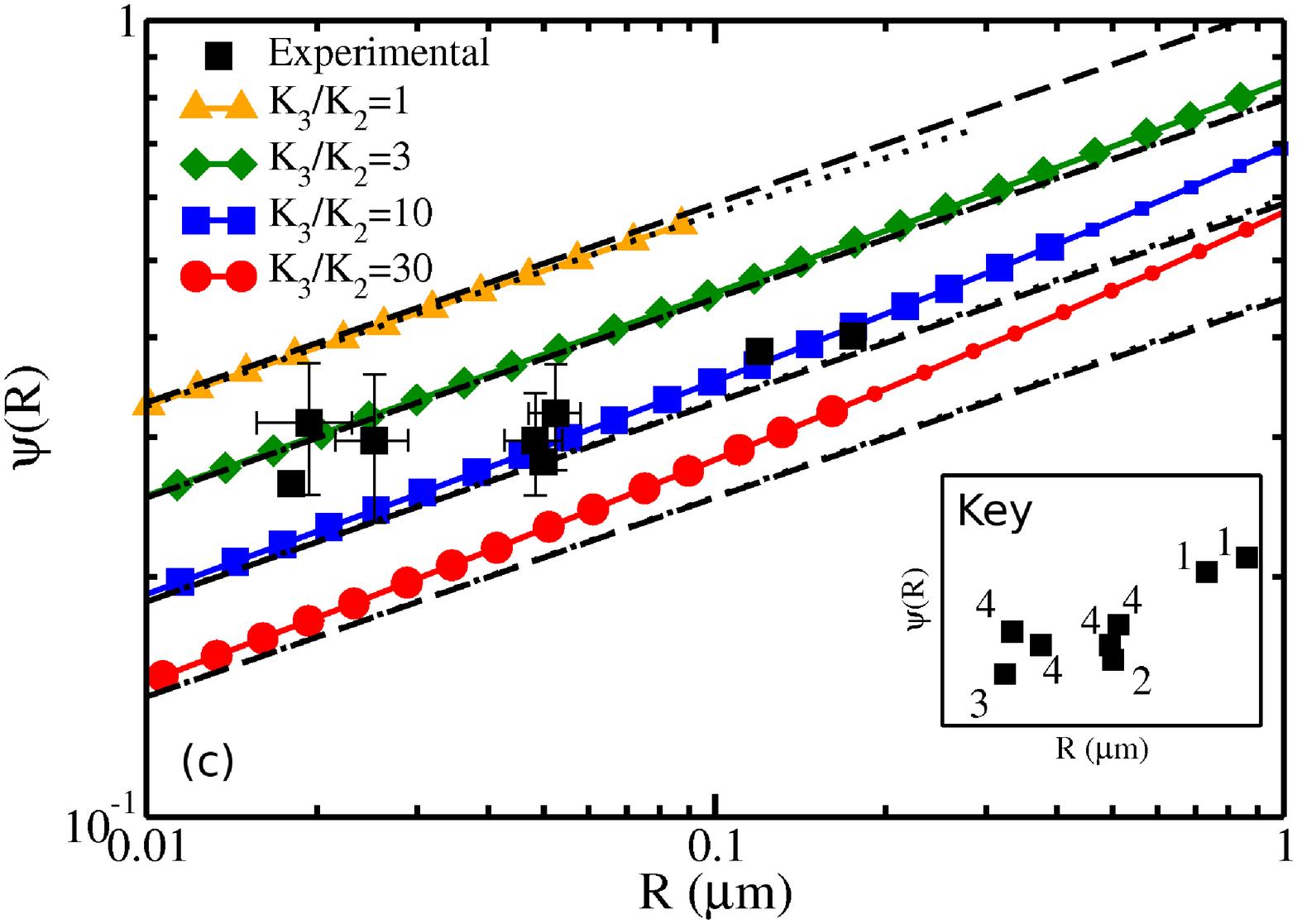} & \includegraphics[width=3.0in]{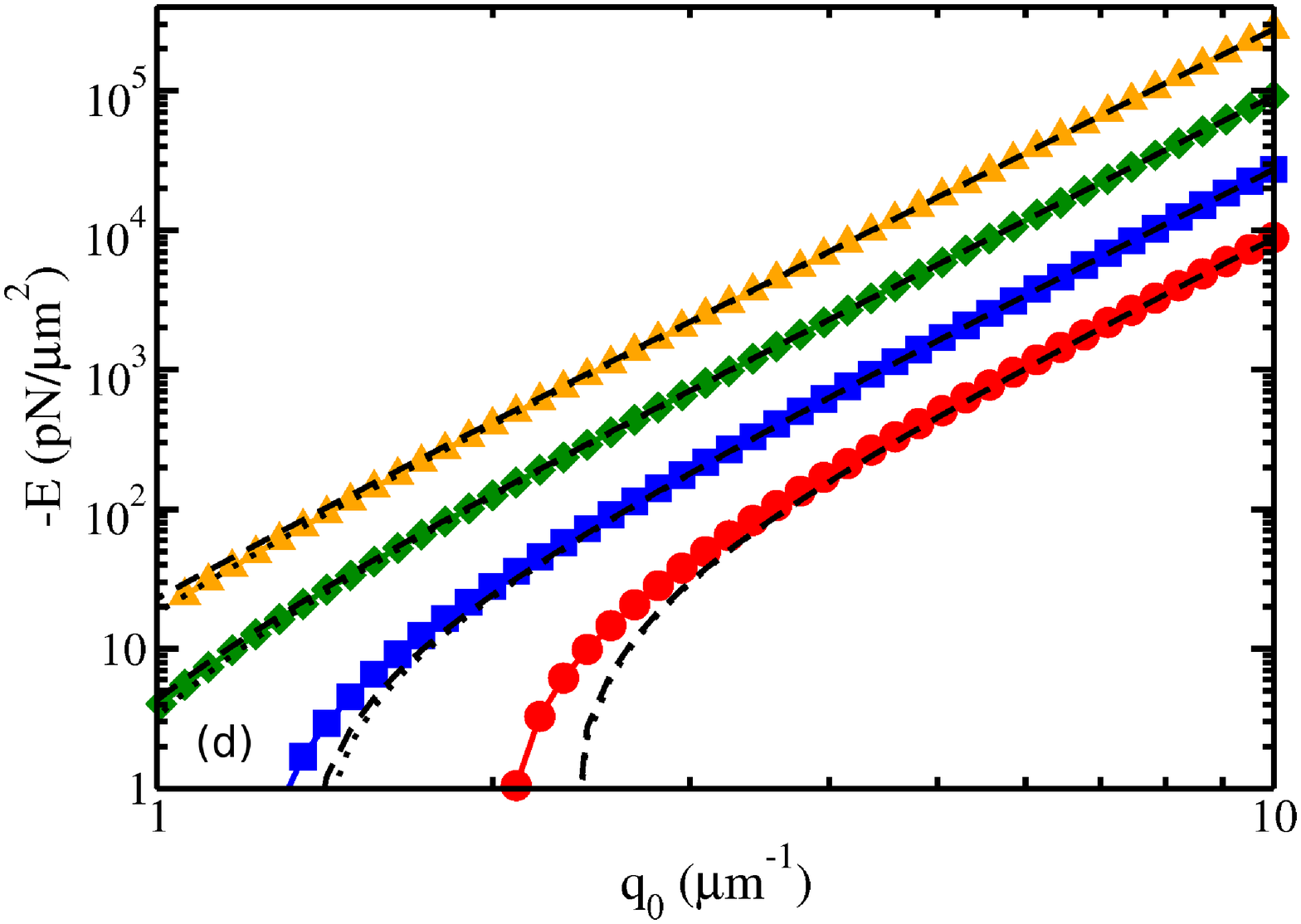}\\
    \end{tabular}
  \caption{Variation of inverse cholesteric pitch $q_0$ for four $K_3/K_2$ ratios as indicated by the legend in (c). Small point sizes indicate positive (unstable) energies.  (a) Radius of equilibrium fibrils $R$ vs.\ inverse cholesteric pitch $q_0$. (b) Equilibrium twist-angle at the surface $\psi(R)$ vs.\ inverse cholesteric pitch $q_0$.  (c) Parametric plot of $\psi(R)$ vs.\ $R$ as $q_0$ is varied.  Experimental data is also shown with black squares. Corresponding to the numbers in the key, and as described in the text, the data comes from Mosser \emph{et al} \cite{mosser06} (key 1), Bouligand \emph{et al} \cite{bouligand85} (key 2), Holmes \emph{et al} \cite{holmes01} (key 3), and Raspanti \emph{et al} \cite{raspanti89} (key 4). (d) Negative of energy per unit volume, $E$, vs.\ inverse cholesteric pitch $q_0$. Positive energies, that are unstable with respect to the cholesteric, are not shown. In all subfigures, dashed black lines are from the small-angle linear approximation, Eqns.~\ref{eq:linearr} - \ref{eq:linearenergy}, while the dotted black lines are from the numerical solution to the linear approximation. We use default parameters $K_2=10$pN, $K_{24}=10$pN, and $\gamma=3$pN/$\mu$m, and all plots are log-log. The shaded grey regions in (a) and (b) indicate the extent of experimental measurements of collagen fibril radius \cite{mosser06, bouligand85, holmes01, raspanti89}.}
  \label{fig:figure3}
\end{figure}

In Fig.~\ref{fig:figure3} we systematically explore the inverse cholesteric pitch $q_0$, for several values of the bend modulus $K_3$ as indicated by the legend in Fig.~\ref{fig:figure3}(c). We explore $q_0$ in the range \cite{mosser06} 1-10$\mu$m$^{-1}$, with $K_3/K_2=1$ \cite{chaikin95, degennes95} to $K_3/K_2 =30$  \cite{lee90, odijk86}.

Fig.~\ref{fig:figure3}(a) shows the equilibrium fibril radius $R$ vs.\ inverse cholesteric pitch $q_0$. $R$ appears to follow a power law of $q_0$ for all $K_3/K_2$ ratios, with an apparent exponent of $-4$. With variation of $q_0$, the data for all $K_3$ values crosses the shaded region showing the range of experimental measurements. The dashed black lines are from the small-angle linear approximation in Eqn.~\ref{eq:linearr}, and show remarkable agreement. In particular, the small-angle linear approximation from Eqn.~\ref{eq:linearr} recovers the observed $R \sim q_0^{-4}$ scaling.  It also captures the approximately linear increase of $R$ as $K_3$ is increased. 

Fig.~\ref{fig:figure3}(b) shows the twist-angle at the surface $\psi(R)$ as a function of the inverse cholesteric pitch $q_0$.  The curves for different $K_3$ values are similar, and all cross the shaded region showing the range of experimental measurements. An approximate power law is seen, and reasonable agreement with the dashed line given by the small-angle linear approximation from Eqn.~\ref{eq:linearpsi}, with $\psi \sim q_0^{-1}$, is seen. Nevertheless the small-angle linear approximation has no $K_3$ dependence, and significant deviations are seen at smaller $q_0$ with larger $K_3$ values. The best agreement is for $K_3/K_2=1$, where the self-consistent stability of the linear approximation holds (with $K_3/K_2 < 2$, see Sec.~\ref{subsec:linear}).  For larger values of $K_3/K_2$ the small-angle linear approximation is no longer self-consistent and we see the effects in the twist-angle at the surface, $\psi(R)$. The linear approximation alone (black dotted curves in Fig.~\ref{fig:figure3}) are not significantly different than the small-angle linear approximation results.

Fig.~\ref{fig:figure3}(c) parametrically plots the twist-angle at the surface $\psi(R)$ against the radius $R$ as $q_0$ is varied. $\psi(R)$ follows an approximate power-law vs.\ $R$, with $\psi \sim R^{1/4}$, as expected from the $q_0$ dependence of Eqns.~\ref{eq:linearr} and \ref{eq:linearpsi}. The deviations from the small-angle linear approximation at larger $R$ are due to the deviation of $\psi(R)$ from the linear approximation for small values of the inverse cholesteric pitch  $q_0$, as we saw in Figs.~\ref{fig:figure3}(a) and (b). The linear approximation (dotted black lines) does not significantly improve upon the small-angle linear approximation results (dashed black lines) --- as expected due to the small surface twist-angles involved. 

The black squares in Fig.~\ref{fig:figure3}(c) show experimental data, where the inset provides a numerical key indicating the source for each data point.  Mosser \emph{et al} \cite{mosser06} (key 1) grew collagen fibrils \emph{in vitro} from a solution from rat tail tendon. From their Fig.~6 we extracted a radius of 175 nm with a twist-angle at the surface of 23$^{\circ}$ and a radius of 120 nm with a surface angle of 22$^{\circ}$. Bouligand \emph{et al} \cite{bouligand85} (key 2) grew collagen fibrils \emph{in vitro} from a solution of calf skin and from their Fig.~13 we extracted a radius of 50 nm and a twist-angle at the surface of 16$^{\circ}$. Holmes \emph{et al} \cite{holmes01} (key 3) used collagen fibrils from adult bovine corneas to measure a fibril radius of 18 nm with a surface twist-angle of 15$^{\circ}$. Raspanti \emph{et al} \cite{raspanti89} (key 4) found a radius of (19.35$\pm$3.7) nm with a twist-angle at the surface of 17.9$^{\circ}\pm3.4^{\circ}$ for collagen fibrils from 6-day-old rat skin, a radius of (105$\pm$10.9) nm with a twist-angle at the surface of $18.4^{\circ}\pm2.8^{\circ}$ from 16-week-old rat skin, a radius of (25.2$\pm$3.7) nm with a twist-angle at the surface of $17.0^{\circ}\pm3.6^{\circ}$ for bovine aorta, and a radius of (48.35$\pm$5.6) nm with a twist-angle at the surface of $17.0^{\circ}\pm1.25^{\circ}$ for bovine optic nerve sheath. The equilibrium model data in Fig.~\ref{fig:figure3}(c) is able to cover the entire range of experimental data by variation of both $q_0$ and $K_3$.  Neither $K_3$ nor $q_0$ alone can explain the experimental variation, but we anticipate significant parameter variation due to the widely varying experimental conditions involved.  Other parameters are explored below.

In Fig.~\ref{fig:figure3}(d) the energies per unit volume are plotted as $q_0$ is varied for different $K_3$ values. On this log-log plot, only negative energies that are stable with respect to the bulk cholesteric are shown. The two lower $K_3$ values have stable (negative) energies for the entire $q_0$ range plotted, while the two higher $K_3$ values are negative for most of the range but are positive at smaller $q_0$.  The energies from the small-angle linear approximation, Eqn.~\ref{eq:linearenergy}, are plotted as dashed lines and begin to significantly disagree with the numerical results for larger values of $K_3/K_2$.  Nevertheless, at larger (negative) energies the approximate scaling of $-E \sim q_0^4$ from the small-angle linear approximation (Eqn.~\ref{eq:linearenergy}) is observed. 

\subsubsection{Surface tension $\gamma$ dependence\\}
\begin{figure}[ht]
\centering
    \begin{tabular}{cc}
      \includegraphics[width=3.0in]{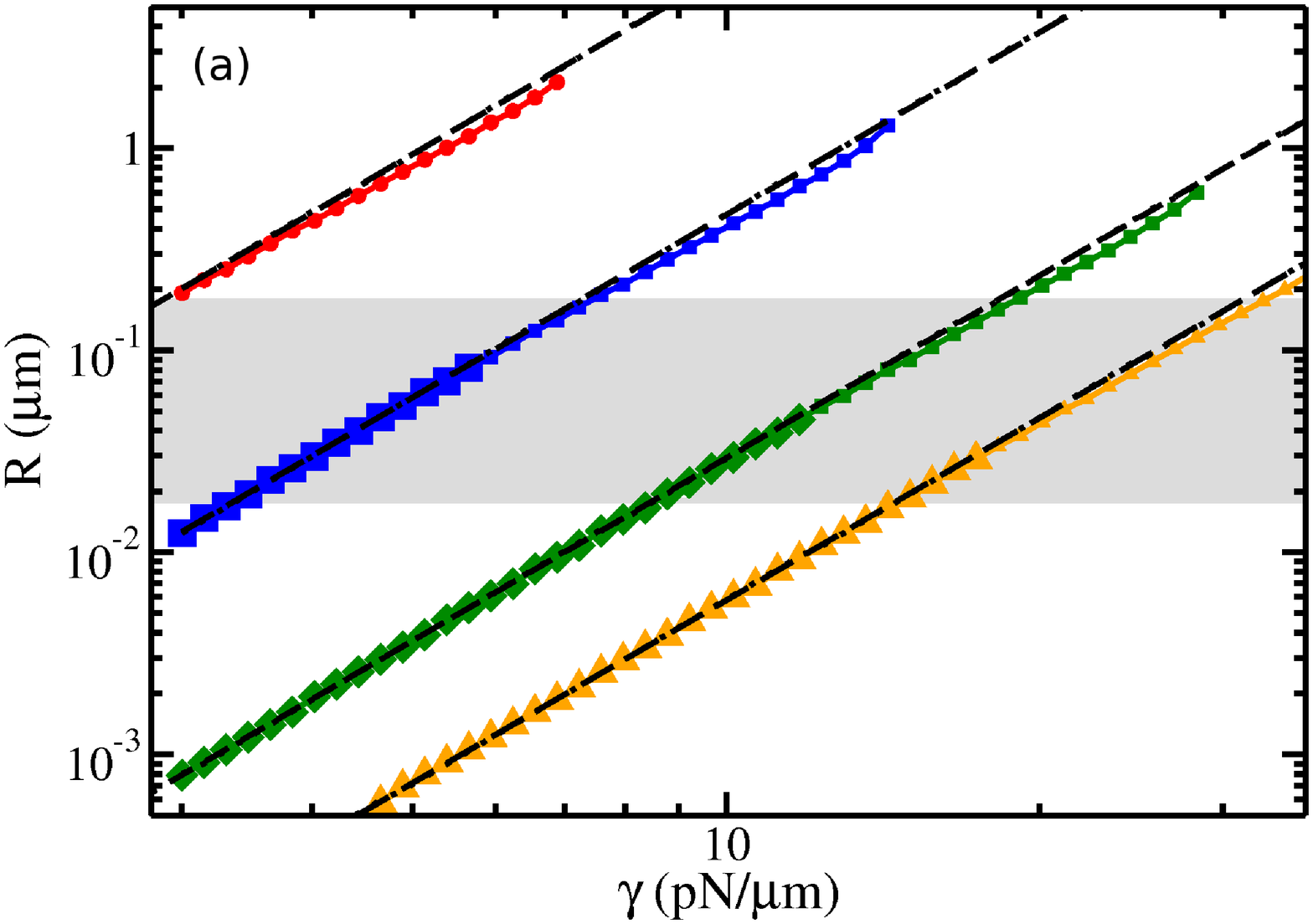} & \includegraphics[width=3.0in]{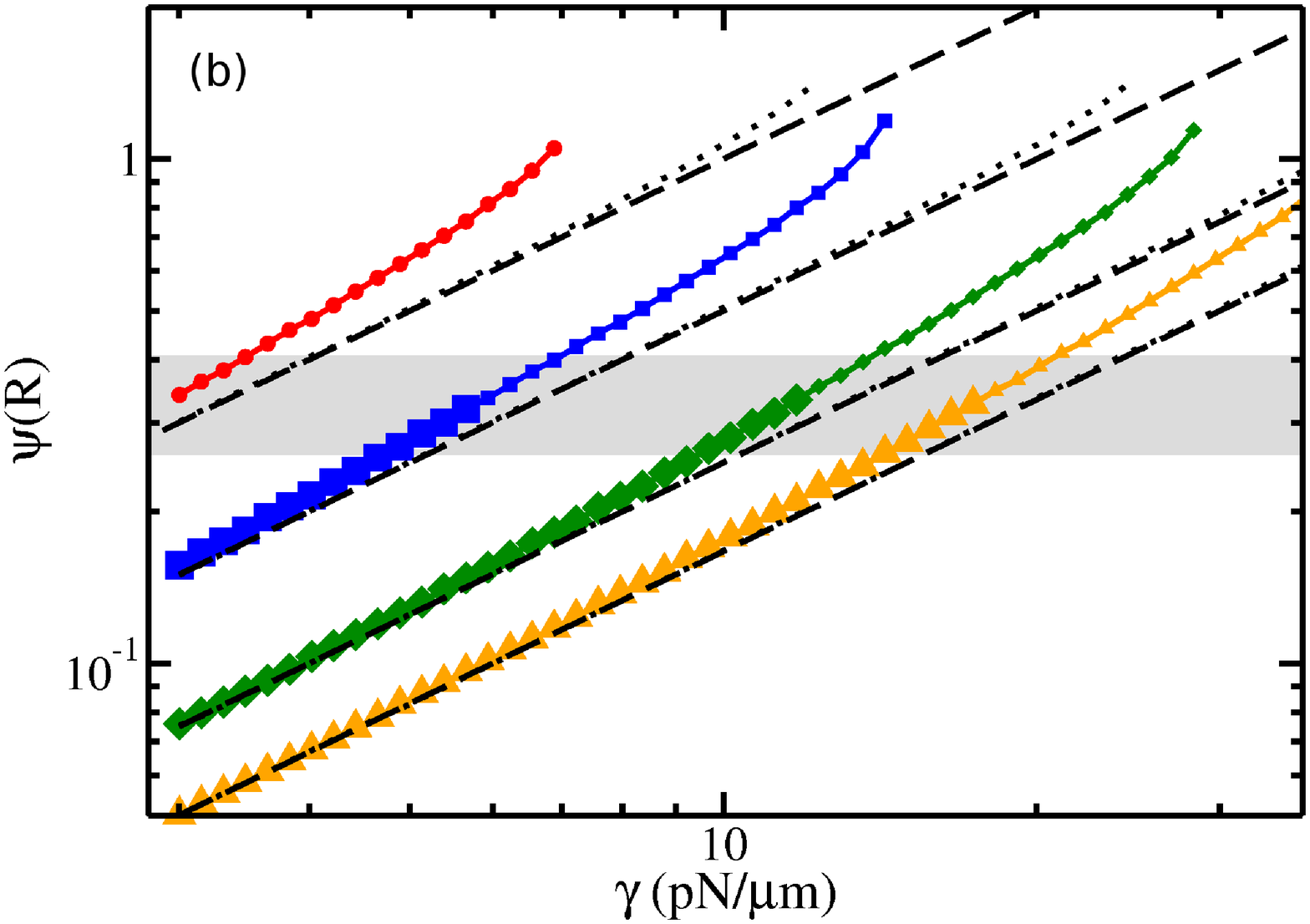}\\
      \includegraphics[width=3.0in]{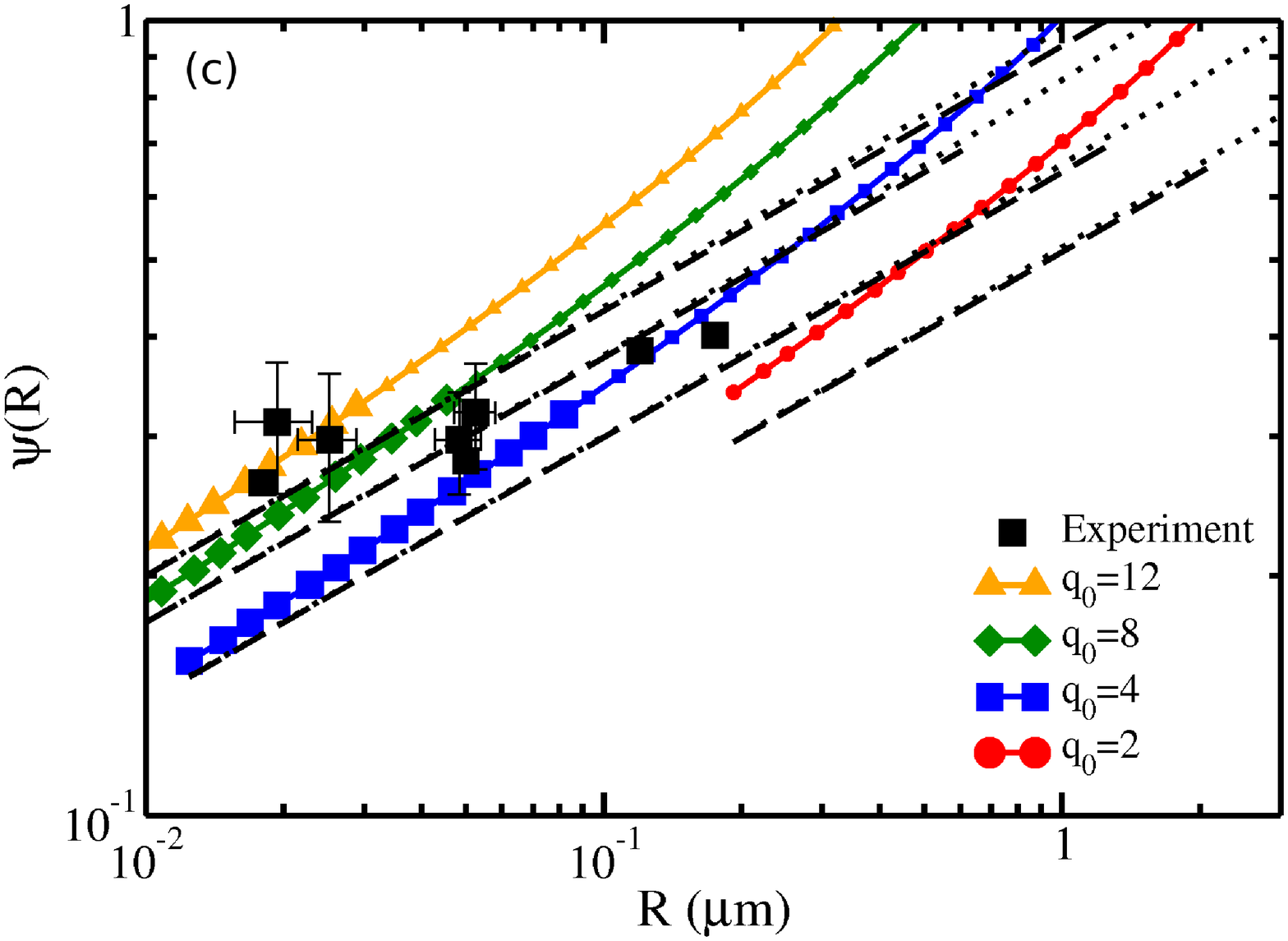} &\includegraphics[width=3.0in]{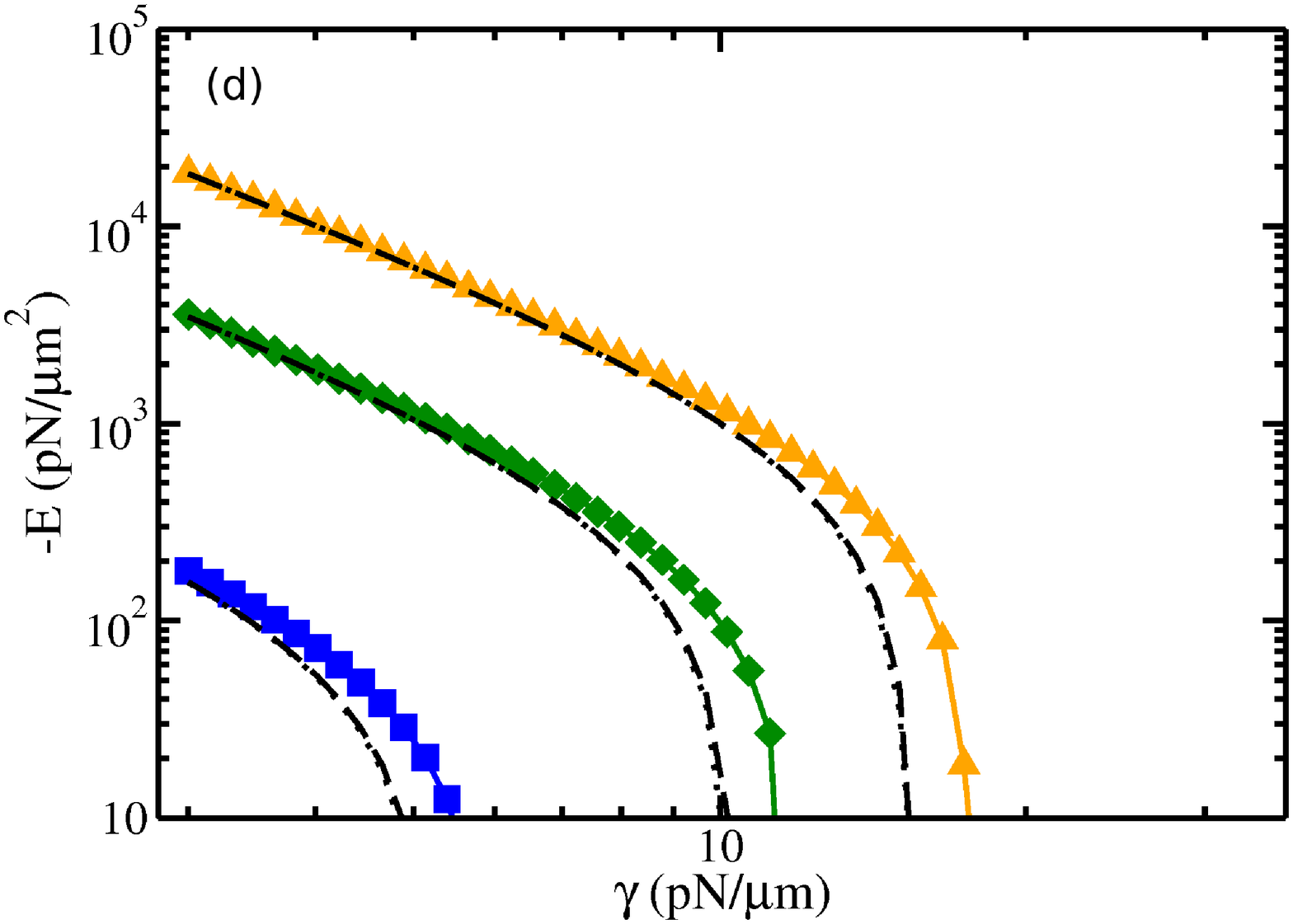}\\
    \end{tabular}
  \caption{Variation of surface tension $\gamma$ for four $q_0$ values as indicated by legend in (c). Small point sizes indicate positive (unstable with respect to the cholesteric) energies. (a) Radius $R$ of equilibrium fibrils vs.\ surface tension $\gamma$.  (b) Equilibrium twist-angle at the surface $\psi(R)$ vs.\  surface tension $\gamma$.  (c) Parametric plot of $\psi(R)$ vs.\ $R$ as $\gamma$ is varied. Experimental data is shown with black squares, with a corresponding key to the sources in Fig.~\ref{fig:figure3}(c). (d) Negative of energy per unit volume, $E$, vs.\ the surface tension $\gamma$. Positive energies, that are unstable with respect to the cholesteric, are not shown. For all subfigures,  dashed black lines are from the small-angle linear approximation, Eqns.~\ref{eq:linearr} - \ref{eq:linearenergy} and dotted black lines are from the numerical solution of the linear approximation. We use default parameter values $K_2=10$pN, $K_3=300$pN, and $K_{24}=10$pN, and all plots are log-log. The shaded grey regions in (a) and (b) indicate the extent of experimental measurements of collagen fibril radius \cite{mosser06, bouligand85, holmes01, raspanti89}.  }
  \label{fig:figure4}
\end{figure}

In Fig.~\ref{fig:figure4} we vary the surface tension $\gamma$ between 3-30 pN/$\mu$m for different values of the inverse cholesteric pitch $q_0$, as indicated in the legend in Fig.~\ref{fig:figure4}(c). Fig.~\ref{fig:figure4}(a) shows the equilibrium fibril radius $R$ vs.\ $\gamma$. We see that $R \sim \gamma^3$, which is in good agreement with the small-angle linear approximation of Eqn.~\ref{eq:linearr} (dashed lines). For the fibril radius, the linear approximation is similar to the small-angle linear and both recover the full model results. Qualitatively, larger $\gamma$ values increase the equilibrium radius since the relative contribution of surface to volume decreases with radius. The shaded region of Fig.~\ref{fig:figure4}(a) shows the range of experimental measurement of fibril radii \cite{mosser06, bouligand85, holmes01, raspanti89}.

Fig.~\ref{fig:figure4}(b) shows the twist-angle at the surface $\psi(R)$ vs.\ $\gamma$. As expected, since larger radii lead to larger angles, the twist-angle at the surface increases with $\gamma$.  The small-angle linear  approximation of Eqn.~\ref{eq:linearpsi}, indicated by dashed lines, gives an approximate scaling of $\psi \sim \gamma$, but detailed agreement is only good at smaller values of $\gamma$ and larger values of $q_0$. The linear approximation alone shows a similar disagreement. While some failure of the linear approximation is expected for the larger (default) value of  $K_3/K_2=30$ appropriate for long tropocollagen complexes, we see that there is still good agreement for smaller $\gamma$ values. The shaded region of \ref{fig:figure4}(b) shows the experimental range of fibril surface angles. 

Fig.~\ref{fig:figure4}(c) parametrically plots the twist-angle at the surface $\psi(R)$ against the radius $R$ as $\gamma$ is varied. Variation of both $\gamma$ and $q_0$ is mostly able to cover the experimental data but, as with the parameter variation in Fig.~\ref{fig:figure3}(c), a single curve is unable to recover all the experimental data. The small-angle linear approximation, as well as the linear approximation, are only good descriptions of the data at smaller $R$ and $\psi$. There, the approximate scaling of the small-angle linear approximation from Eqns.~\ref{eq:linearr} and \ref{eq:linearpsi}, with $\psi \sim R^{1/3}$ as $\gamma$ is varied, is observed. 

In Fig.~\ref{fig:figure4}(d) the equilibrium energies per unit volume are plotted as $\gamma$. Only stable (negative) energies with respect to a bulk cholesteric phase are shown. As with the small-angle linear approximation in Eqn.~\ref{eq:linearenergy}, we require smaller values of $\gamma$ and/or smaller values of $q_0$ for stability. The energy curves from the linear small-angle approximation, Eqn.~\ref{eq:linearenergy} (black dashed curves), as well as the linear approximation (black dotted curves), are qualitatively similar to our full model results but differ significantly as the energy decreases towards zero.

\subsubsection{Saddle-splay modulus $K_{24}$ dependence\\}
\begin{figure}[ht]
\centering
    \begin{tabular}{cc}
      \includegraphics[width=3.0in]{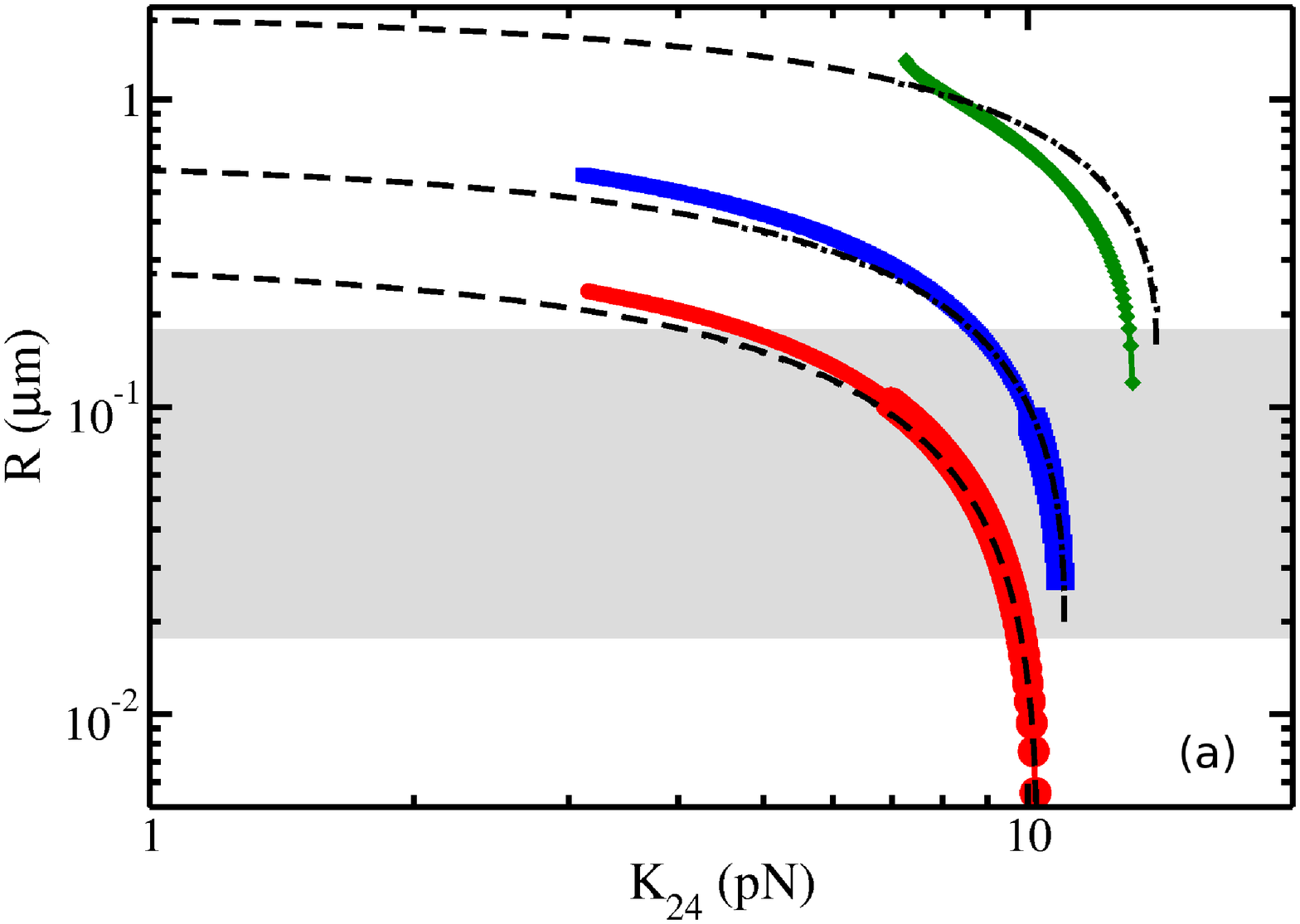} & \includegraphics[width=3.0in]{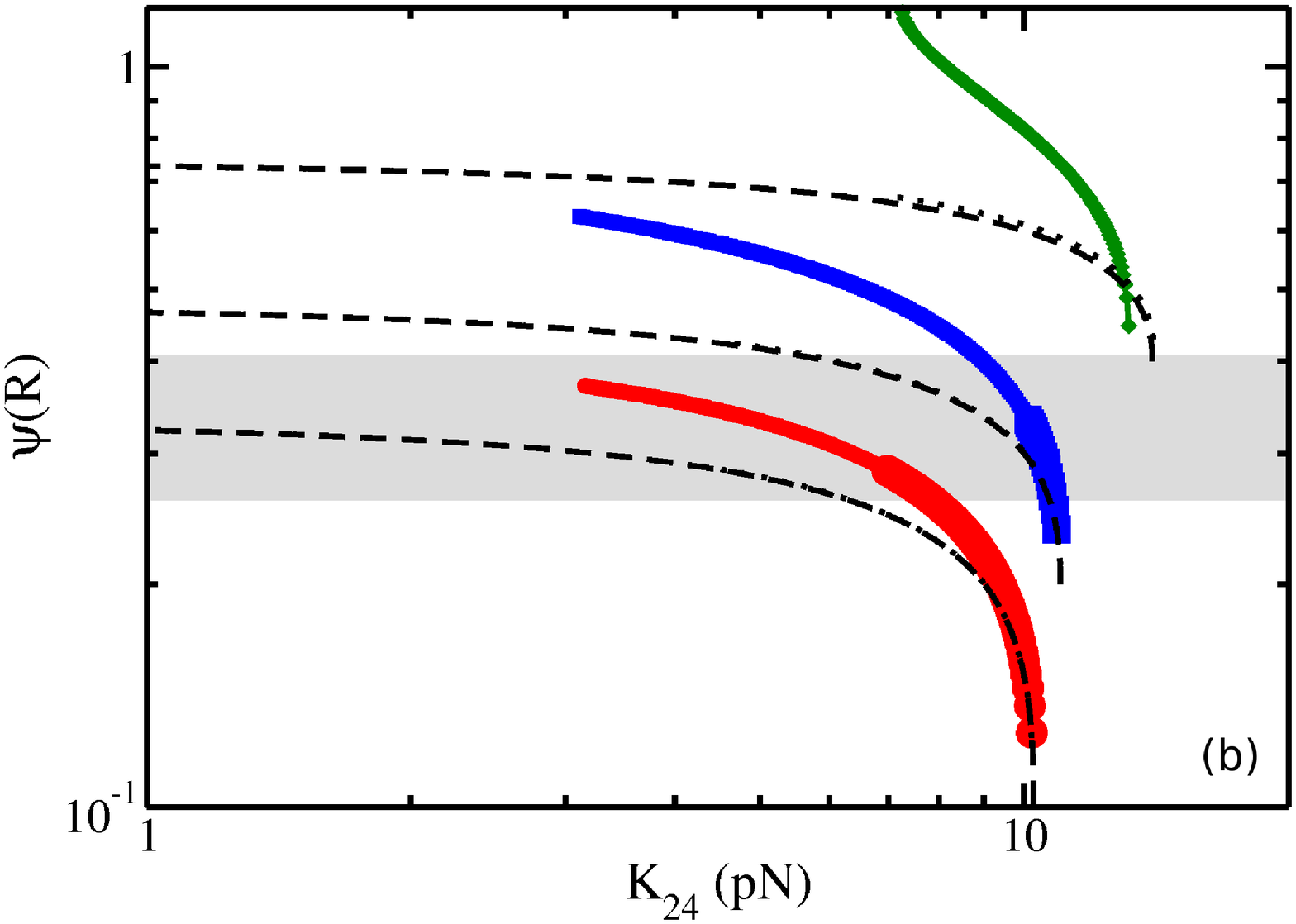}\\
      \includegraphics[width=3.0in]{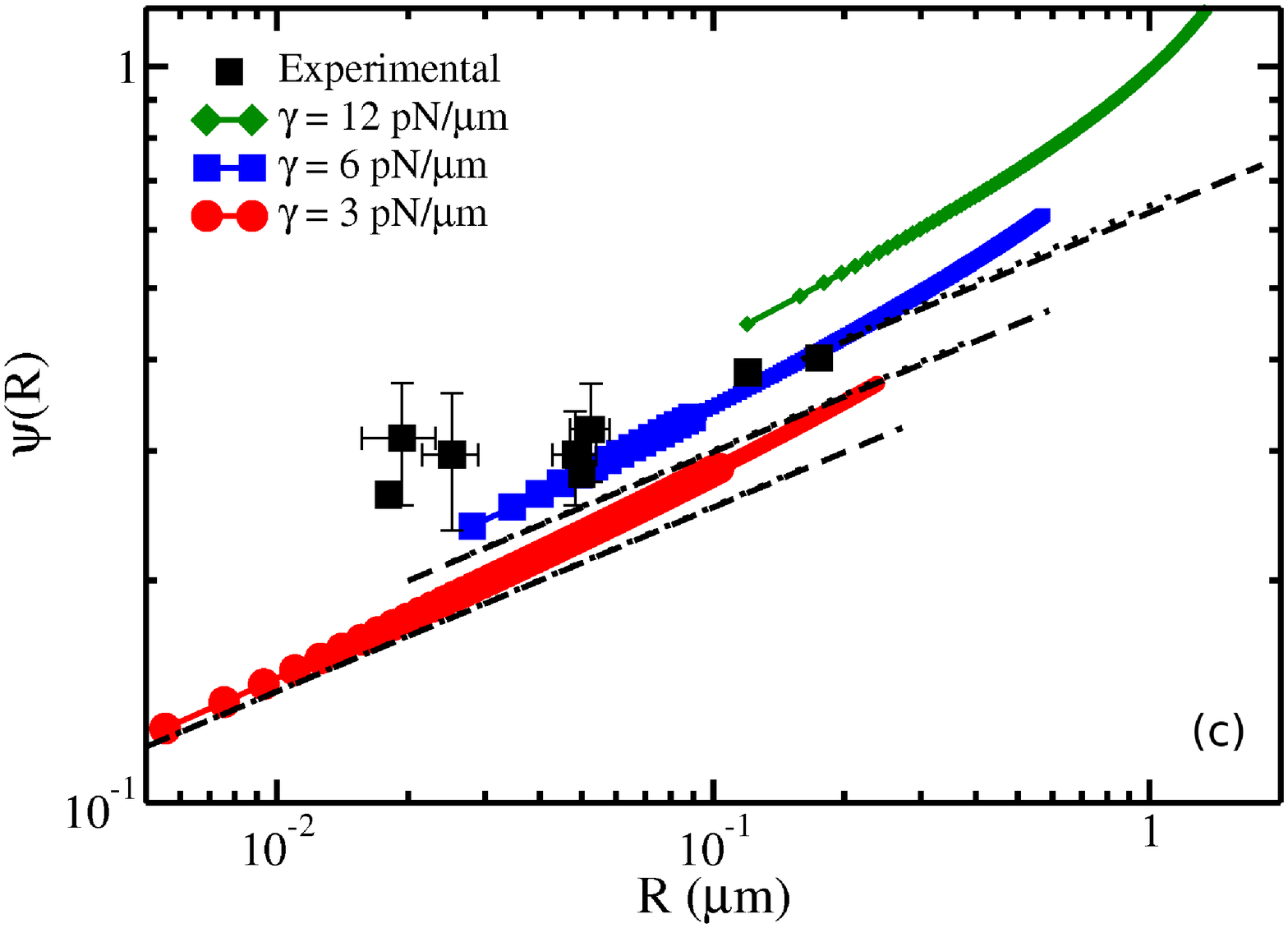} &\includegraphics[width=3.0in]{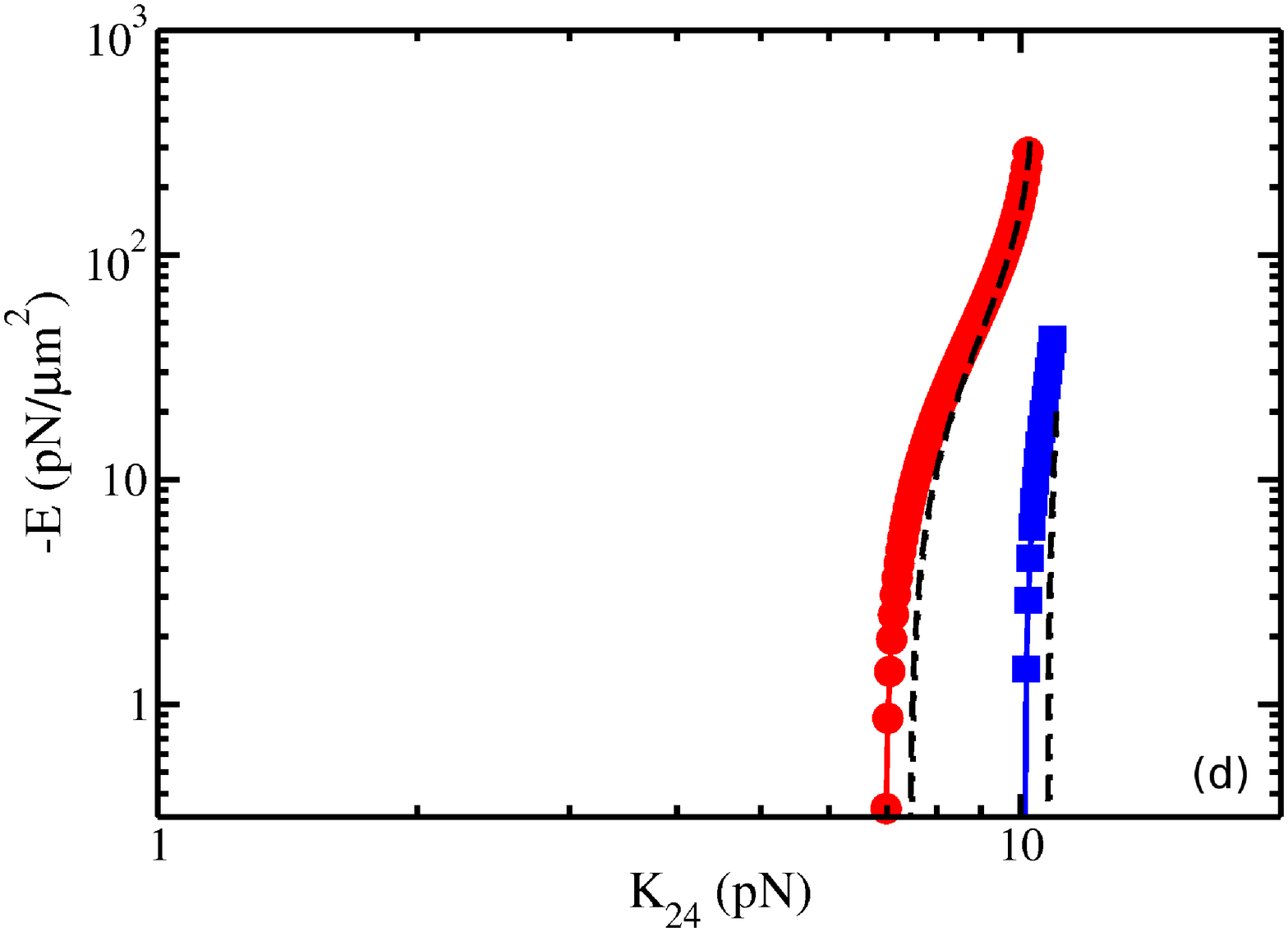}\\
    \end{tabular}
  \caption{Variation of saddle-splay modulus $K_{24}$ for three $\gamma$ values as indicated by legend in (c). Smaller point sizes indicate positive (unstable) energies.  (a) Radius $R$ of equilibrium fibrils vs.\ $K_{24}$.  (b) Equilibrium twist-angle at the surface $\psi(R)$ vs.\  $K_{24}$.  (c) Parametric plot of $\psi(R)$ vs.\ $R$ as $K_{24}$ is varied. Experimental data is shown with black squares, with a corresponding key to the sources in Fig.~\ref{fig:figure3}(c). (d) Negative of the energy per unit volume, $E$, vs.\ $K_{24}$. Positive energies, that are unstable with respect to the cholesteric, are not shown. For all subfigures,  dashed black lines are from the small-angle linear approximation, Eqns.~\ref{eq:linearr} - \ref{eq:linearenergy} and dotted black lines are from the numerical solution of the linear approximation. We use $q_0=4\mu$m$^{-1}$ and otherwise use default parameter values $K_2=10$pN, $K_3=300$pN, and $\gamma=3$pN/$\mu$m, and all plots are log-log.  The shaded regions indicate the extent of experimental measurements of collagen fibril radius \cite{mosser06, bouligand85, holmes01, raspanti89}.
  }
  \label{fig:figure5}
\end{figure}

In Fig.~\ref{fig:figure5} we vary the saddle-splay constant $K_{24}$ for several values of the surface tension $\gamma$. While \cite{meiboom83} $K_{24} = \frac{1}{2}(K_1 + K_2)$ , the splay modulus $K_1$ has no direct effect on the energetics of our divergence-free double-twist fibril structure. We can think of $K_{24}$ variation as implicitly varying $K_1 = 2 K_{24}-K_2$, with $K_{24}=K_2$ corresponding to $K_1=K_2=K_{24}$. Fig.~\ref{fig:figure5}(a) shows the fibril radius $R$ vs.\ $K_{24}$. For $K_{24} \gtrsim K_2 =10$pN the radius $R$ decreases significantly, with sharper decreases at larger values of $K_{24}$ for larger $\gamma$ values.  Fig.~\ref{fig:figure5}(b) has the twist-angle at the surface $\psi(R)$ vs.\ $K_{24}$. For $K_{24} \gtrsim K_2 = 10$pN, we  observe a  similar sharp decrease of $\psi$ as $K_{24}$ increases. 

Using the small-angle linear approximation of section \ref{subsec:linear}, Eqns.~\ref{eq:linearenergyfirst} and \ref{eq:linearrfirst} lead to a cubic equation for $\psi'_{lin}$ for the general case when $K_{24} \neq K_2$. Numerically solving the cubic  \cite{press07} for $\psi'_{lin}$ leads to the dashed black lines in Figs.~\ref{fig:figure5} (a) - (d). The small-angle linear solution agrees well with the full model results at lower surface tensions $\gamma$, but loses accuracy as the surface tension increases. The agreement is better for $R$ than for $\psi(R)$. The linear approximation without the small-angle approximation, shown with dotted lines in Fig.~\ref{fig:figure5}, differs little from the small-angle linear approximation. 

Fig.~\ref{fig:figure5}(c) parametrically plots the twist-angle at the surface $\psi(R)$ against the radius $R$ as $K_{24}$ is varied. Variation of $\gamma$ and $K_{24}$ is able to recover some of the experimental data points, shown with black squares. The black dashed curves from the small-angle linear approximation of section \ref{subsec:linear} have a power law of 1/4, which follows from Eqn.~\ref{eq:linearrfirst} combined with the linear approximation $\psi(R)=\psi'_{lin}R$. The model data appears to approach this power law for lower $R$. 

In Fig.~\ref{fig:figure5}(d) the energies per unit volume are plotted as $K_{24}$ is varied for different $\gamma$ values. The highest $\gamma$ value is not shown as it does not have any negative energies. The magnitude of the energies increases as $\gamma$ decreases and as $K_{24}$ increases. As $K_{24}$ decreases, the energy at each $q_0$ value eventually becomes positive, setting a lower limit on $K_{24}$ for stability. Only a narrow range of $K_{24}$, close to $K_2$, appears to give stable double-twist fibrils.

\subsection{Alternative radial fibril structure models}
With our model free-energy, it is straight-forward to address the energetics of a proposed axial core with constant twist-angle sheath model of collagen fibrils \cite{raspanti11}.  For a core-radius $R_C$ and fibril radius $R$, we evaluate the Frank free energy Eqn.~\ref{eq:frankparameter} using an axial core with $\psi(r)=0$ for $0<r<R_C$ together with a constant twist-angle sheath with  $\psi(r)=\psi_0$ for $R_C<r<R$.  In the core, the free energy density is $f_C=K_2q_0^2/2$. In the sheath, the free energy density is $f_0=(K_2/2)[q_0-(1/r)\sin\psi_0\cos\psi_0]^2+(K_3/2)(\sin^4\psi_0/r^2)$. Significantly, the saddle-splay term does not contribute since it has equal and opposite contributions at the surfaces at $R_C$ and at $R$. Since $f_C$ and $f_0$ and the surface tension contributions are all positive, the total free energy will also be positive for all values of $R$, $R_C$, and $\psi_0$ --- above the energy of the bulk cholesteric phase. The same argument and conclusion also applies to a constant twist-angle collagen fibril model, where $R_C=0$. Neither the axial core constant twist-angle sheath model nor the constant twist-angle model for collagen fibrils are stable equilibrium phases with respect to the bulk cholesteric, or with respect to the double-twist fibrils presented in this paper. 

\section{Discussion}
At high concentrations collagen forms a cholesteric phase with pitch between 0.5 and 2 $\mu$m\cite{mosser06}, while at lower concentrations tropocollagen complexes aggregate into packed fibrils \cite{hulmes02}. Fibril radius depends on assembly conditions \cite{hulmes02}. Tropocollagen complexes are tilted with respect the fibril axis \cite{doucet11, holmes01, raspanti89, lillie77, bouligand85}, which is consistent with a double-twist structure for collagen fibrils \cite{bouligand85}.  

We propose that an equilibrium double-twist structure for collagen fibrils is determined by a liquid-crystalline Frank free-energy similar to that used for the blue phases of liquid crystals \cite{degennes95, meiboom81, meiboom83, wright89}.  We use an Euler-Lagrange approach to minimize the free-energy for a fibril of a given radius $R$, and then numerically minimize the free-energy per unit volume as a function of fibril radius as illustrated in Fig.~\ref{fig:figure1}(a). The existence of a clear minimum as a function of $R$ implies that the uniform fibril radius observed both \emph{in vivo} and \emph{in vitro} may be selected by equilibrium free energy considerations. Our results show how this minimum determines both the fibril radius and the twist-angle at the surface as system parameters are varied. 

In studies of blue-phases, the director twist-angle is typically assumed to be a linear function of radius \cite{degennes95, meiboom81, meiboom83, wright89}. Although the linear approximation is good close to the fibril core, Fig.~\ref{fig:figure1}(d) shows the twist-angle can significantly deviate from linearity near the fibril surface.  Nevertheless, with a small-angle assumption and for equal twist and saddle-splay moduli ($K_2=K_{24}$), the linear approximation gives an analytical minimal free-energy solution with power-law relationships between fibril radius $R$ and twist-angle at the surface $\psi(R)$ vs.\ the system parameters. As shown in Figs.~\ref{fig:figure3}-\ref{fig:figure5}, the resulting power-law scaling is a good approximation for most of the parameter ranges and provides a useful guide for understanding the relationship between radius and twist-angle at the surface. We have also evaluated the linear approximation numerically and found little difference with the small-angle linear approximation, indicating that disagreements with our full model results are due to the linear approximation alone. This is consistent with the relatively small twist-angles in our results.  We find that the linear approximation is less accurate at higher bend modulus ($K_3$), lower inverse cholesteric pitch ($q_0$), or higher surface tension ($\gamma$). 

We expect that the bend modulus, $K_3$, is large for collagen. While the commonly used equal modulus approximation \cite{chaikin95, degennes95} assumes that $K_3=K_2$, we know that molecules with large aspect ratio can have $K_3$ as large as $30K_2$ \cite{lee90, odijk86}.  Fig.~\ref{fig:figure3} shows that $K_3 > K_2$ appears to be necessary for our model to be consistent with the experimental collagen fibril radii and surface-twist angles. Larger $K_3$ values also lead to narrower energy wells, as illustrated in Fig.~\ref{fig:figure2}.

We can recover \emph{in vitro} and \emph{in vivo} experimental \cite{mosser06, bouligand85, holmes01, raspanti89} values of fibril radius $R$ and twist-angle at the surface $\psi(R)$ with reasonable variation of many of our model parameters.  Unfortunately, the twist-angle at the surface and radius alone are not sufficient to determine all of the model parameters --- so explicit experimental control or measurement of model parameters will be needed to assess our equilibrium model of radius and twist-angle control in collagen fibrils.  Here, we emphasize $q_0$ and $\gamma$ since they appear to be the most experimentally accessible parameters in the collagen system, and we predict a robust power law dependence of fibril radius and twist-angle at the fibril surface for these two parameters.

As shown in Fig.~\ref{fig:figure3}, our model is consistent with experimental results only over a roughly two-fold variation of the inverse cholesteric pitch, $q_0$. These $q_0$ values are close to those expected from the observed cholesteric collagen pitch \cite{mosser06}. Qualitatively our model predicts an increase in radius and surface twist-angle as $q_0$ is decreased, which is consistent with increased surface-twist angles in fibrils swollen with urea solution \cite{lillie77}.  Nevertheless, $q_0$ is difficult to vary by large factors. Even over quite variable \emph{in vitro} and \emph{in vivo} conditions \cite{livolant91}, the pitch of DNA only varied by a factor of 10, and only over a factor of 2 using concentration and ionic strength \cite{stanley05}.

The surface tension is investigated in Fig.~\ref{fig:figure4}. Higher surface tensions lead to larger radii and twist-angles at the surface. Variation of surface tension moves the model results through the experimental measurements of surface twist-angle as a function of radius. We expect that surface tension, reflecting surface energy of the fibril, could be modified by surfactants, by surface modifications of collagen fibrils, and by the environment surrounding collagen fibrils. The range of surface tensions that agree with experimental measurements puts limits on the surface tensions of collagen fibrils and similar protein aggregates --- from 3pN/$\mu$m to approximately 20pN/$\mu$m.

Modifications to fibril surfaces would be expected to affect the surface tension $\gamma$ but not the elastic constants. The reported increase of collagen fibril radius in animal models with knockouts of proteoglycans \cite{danielson97, tasheva02} are intriguing in this respect. Proteoglycans decorate collagen fibrils \cite{scott84}, and so would be expected to modify $\gamma$. The reported increase of fibril radius with a decrease of proteoglycan \cite{danielson97, tasheva02, scott84} is consistent, according to Eqn.~\ref{eq:linearr} and Fig.~\ref{fig:figure4}, with proteoglycans acting as effective surfactants that decrease the surface tension $\gamma$.  It would therefore be interesting to measure fibril radius dependence on proteoglycan for {\em in vitro} systems, where the twist-angle at the surface is also assessed. 

The elastic and surface parameters of our model are effective, or coarse-grained, properties of the fibril. We consider only tropocollagen alignment, and not the placement of individual molecules. As such we effectively coarse-grain the well known axial D-banding of collagen fibrils (see e.g. \cite{hulmes02}). As D-bands are not correlated with significant modulations of radius along the fibril axis (see e.g. \cite{holmes01, bouligand85}), this appears to be a reasonable approximation. We expect that mixtures of different types of collagen would lead to different effective parameters that will depend on the mixture. We suspect that this explains the systematic variation of fibril radius and twist-angle at the surface observed with mixtures of collagen I and V \cite{birk90}. We note that \emph{in vivo}, the detailed environment of collagen fibrillogenesis \cite{Banos08} in different tissue types may also significantly change effective parameters. This may be able to explain the particularly small fibril radius observed in the cornea, which variations of fibril composition alone are unable to replicate \cite{birk90}. We note that age-related cross-linking, important for mechanical properties of collagen \cite{Avery08}, will essentially lock in the equilibrium structure even after the microenvironment of the fibril has changed. 

Finally, our equilibrium double-twist model may also apply to other fibrillar systems. For example, human hair orthocortex macrofibrils \cite{harland14, bryson09}, composed of the intermediate filament keratin, appear to have a double-twist structure. This is seen also in wool \cite{mckinnon11, harland11, caldwell05, rogers59}.

\section{Conclusion}
We use a liquid crystal model of collagen fibrils to compare to experimental measurements of collagen fibril radius $R$ and surface twist-angle $\psi(R)$. Using an Euler-Lagrange approach and numerical minimization, we demonstrate the existence of a minimum in free energy as a function of fibril radius $R$, suggesting fibril radius can be determined by the equilibrium free energy. Large bend modulus, $K_3$ in the liquid crystal Frank free energy, and small surface tension $\gamma$ are found to be necessary to agree with experimental measurements. By varying the model parameters, most significantly $K_3$, $\gamma$, and $q_0$, the model is able to recover the same range as observed in experimental measurements. We expect that different tissue environments, collagen type makeup of a fibril, and other interacting proteins will lead to different effective parameters in our model, and allow tissues to vary the characteristics of equilibrium collagen fibrils.

\section{Acknowledgements}
We thank the Natural Science and Engineering Research Council (NSERC) for operating grant support, and the Atlantic Computational Excellence Network (ACEnet) for computational resources. AIB thanks NSERC, the Sumner Foundation, and the Killam Trusts for fellowship support.

\footnotesize{
\bibliography{DoubleTwist} 
\bibliographystyle{rsc} 
}
\end{document}